\documentclass[11pt]{article}

\usepackage{amssymb}
\usepackage{subfig}
\usepackage{color}
\usepackage{epsfig,amssymb,amsfonts,amsmath,graphicx,dsfont,cite,xfrac}
\usepackage{authblk}
\definecolor{mygray}{gray}{0.5}
\usepackage{cite}
\usepackage[colorlinks=true,linkcolor=blue,citecolor=red]{hyperref}
\usepackage{multirow}

\usepackage{tikz,pgfplots}
\usetikzlibrary{decorations.pathreplacing}
\usetikzlibrary{shapes.multipart}
\usetikzlibrary{positioning}
\usetikzlibrary{matrix}
\usetikzlibrary{shapes,calc,arrows}

\newcommand{\be}{\begin{equation}}
\newcommand{\ee}{\end{equation}}
\newcommand{\bea}{\begin{eqnarray}}
\newcommand{\eea}{\end{eqnarray}}

\title{Lieb lattices and pseudospin-1 dynamics under barrier- and well-like electrostatic interactions}

\author[${}$]{V. Jakubsk\'y${}^{1}$ and K. Zelaya${}^{1}$}

\affil[${1}$]{\footnotesize Nuclear Physics Institute, Czech Academy of Science, 250 68 \v{R}e\v{z}, Czech Republic}

\date{}

\begin{document}
	
\maketitle
	
\begin{abstract}
This work considers the confining and scattering phenomena of electrons in a Lieb lattice subjected to the influence of a rectangular electrostatic barrier. In this setup, hopping amplitudes between nearest neighbors in orthogonal directions are considered different, and the next-nearest neighbor interaction describes spin-orbit coupling. This makes it possible to confine electrons and generate bound states, the exact number of which is exactly determined for null momentum parallel to the barrier. In such a case, it is proved that one even and one odd bound state is always generated, and the number of bound states increases for non-null and increasing values of the parallel momentum. That is, current-carrying states are generated. In the scattering regime, the energies are determined so that resonant tunneling occurs. The existence of perfect tunneling energy in the form of super-Klein tunneling is proved to exist regardless of the bang gap opening. Finally, it is shown that perfect reflection appears when solutions are coupled to the intermediate flat-band solution.

\end{abstract}
	
\section{Introduction}
The theoretical and experimental progress in the physics of graphene and other Dirac materials has become a trending topic in material science and theoretical physics~\cite{Weh14,Aok14}. Many remarkable properties of these materials follow from the fact that dynamics of low-energy quasi-particles is described by equations known in relativistic quantum mechanics. It makes it possible to test relativistic properties such as Klein tunneling~\cite{Kat06,All11}, relativistic Landau levels, and the existence of pseudoparticles violating the Lorentz invariance~\cite{Yan17,Zha17} (type-II Dirac fermions). Graphene mono- and multi-layer systems exhibit transport properties such as quantum Hall effect~\cite{Can06} and anomalous quantum Hall effect~\cite{Hal88}, and Josephson effect in twisted cuprate bilayers~\cite{Tum22}.

Graphene has shown to be a helpful benchmark system to test the properties of relativistic pseudospin-$1/2$ particles in low-energy systems. Nevertheless, the family of Dirac materials also contains other, equally interesting, members. Their geometries can extend beyond the honeycomb lattice. For instance, there are Kagome~\cite{Mek03}, Dice or $\alpha-T_{3}$~\cite{Ill17,Dey19}, and Lieb lattices~\cite{Gol11,Jak23}, which lead to effective pseudospin-1 Dirac equations. It was recently showed that the Kagome lattice can be obtained from a geometrical deformation of the Lieb lattice~\cite{Jiang}. For a recent survey of two-dimensional lattices and their physical properties and realization, see~\cite{Fan22}.

Particularly, the Lieb lattice is a two-dimensional array with a periodicity of a square lattice. The sites are located in the corners of each square and at the midpoints on its sides. To our best knowledge, the Lieb lattice has not been found in nature. However, it has been prepared artificially in diverse ways \cite{Yan19,Ley18}, and was realized in experiments with optical fibers \cite{Guz14,Vic14,Die16,Vic15,Muk15}. Furthermore, it was formed by ultracold atoms trapped in optical lattices \cite{Shen10} or by electrons of Cu(111) atoms confined by an array of CO molecules~\cite{Slo17}. It was also prepared in covalent-organic frameworks~\cite{Cui20}. 

The tight-binding model accurrately describes the band structure of the Lieb lattice, which reveals the existence of two bands with positive and negative energies, and an additional so-called flat band. The latter is associated with the states that have fixed  (zero) energy independent of the value of momentum. It is worth mentioning that the flat band solutions were prepared in the optical experiments~\cite{Vic15,Muk15}. Similarly to graphene, the dynamics of the low-energy quasi-particles in the Lieb lattice is dictated by a relativistic Dirac-type equation. Nevertheless, these quasi-particles have pseudospin-1 due to the existence of three atoms per unit cell.

In the current article, we investigate the scattering and confinement of the relativistic quasi-particles by a rectangular electric potential in the Lieb lattice with a gapped band structure. Gap-opening can be induced by on-site energy that differs on three sublattices or by the phase acquired by the electron when jumping between the neighboring sites \cite{Shen10}, see also \cite{Green}. In this article, we adopt the second approach where a purely imaginary next nearest-neighbor interaction, attributed to spin-orbital coupling \cite{Jak23}, is taken into account. 

Effects such as electron confinement and transmission are obtained with the aid of the proper boundary conditions, which enforce the continuity on two out of the three pseudospin-1 components. The third component can be discontinuous, which leads to a spatial discontinuity in the probability density. Nevertheless, it does not compromise the associated continuity equation. Electron dynamics for electrostatic interactions in graphene have been discussed in the literature, such as the transmission properties in square barriers~\cite{Kly08,Tao13} and electron confinement with cylindrical quantum dots~\cite{Bar09}. We thus focus on the related properties of the quasi-particle dynamics in the Lieb lattice. We further analyze the influence of the flat-band solution in electron dynamics. As shown in the manuscript, solutions in this regime are described by degenerate Bloch-wave solutions whose linear combinations can compose wavepackets of arbitrary form. These are shown to be current-free solutions regardless of the nature of the wavepacket. As a result, one obtains perfectly reflected waves when they couple to flat-band solutions.

The manuscript is structured as follows. In Sec.~\ref{sec:Lieb} we briefly introduce and discuss the main properties of the Lieb lattice with nearest and next-nearest neighbor interactions, from which the effective low-energy Dirac equation is obtained. In Sec.~\ref{sec:null-mag}, we present the general solutions and the transfer matrix associated with the rectangular electrostatic interaction. The latter is then exploited in Sec.~\ref{subsec:U-confinement} and Sec.~\ref{subsec:U-scattering} to discuss in full detail the localization of electrons and scattering dynamics, respectively. Finally, discussions and perspectives are provided in Sec.~\ref{sec:conclusions}, and complementary details about the proof of the number of bound states are given in App.~\ref{sec:even-odd}.

\section{Lieb lattice and pseudospin-1 Dirac equation}
\label{sec:Lieb}
\begin{figure}
\centering
\subfloat[][]{\includegraphics[width=0.45\textwidth]{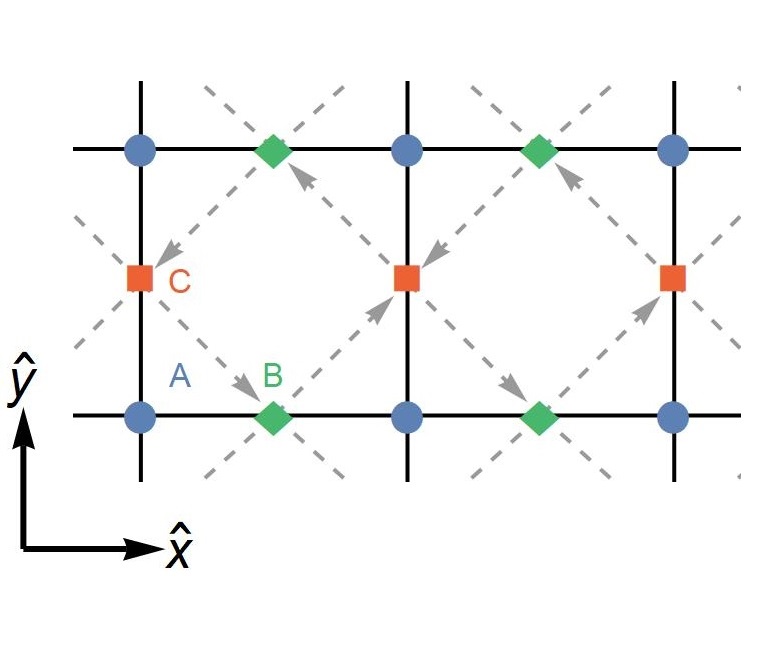}
\label{fig:Lieb-1}}
\hspace{2mm}
\subfloat[][]{\includegraphics[width=0.3\textwidth]{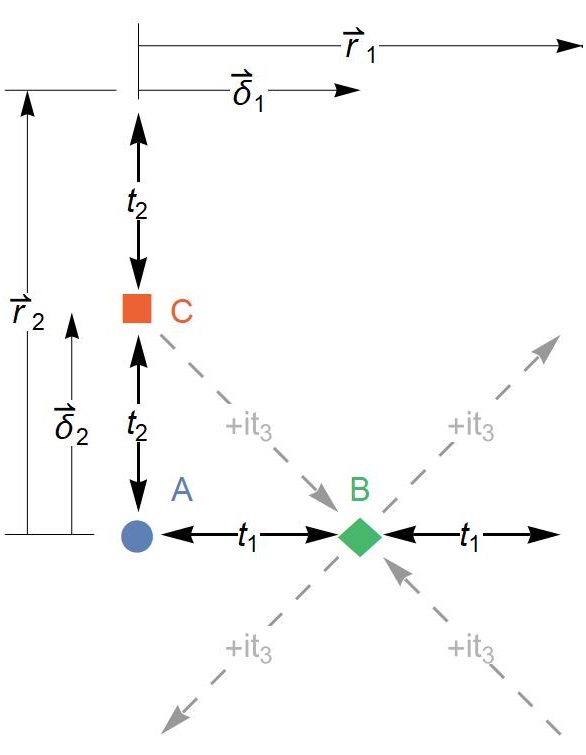}
\label{fig:Lieb-2}}
\caption{(a) Lieb lattice, composed by the atoms $A$ (blue circle), $B$ (green diamond), and $C$ (red-filled square). The dashed arrows denote the direction of positive phase hopping parameter between next-nearest neighbors $B-C$.  (b) Composition of a unit cell of the Lieb lattice. The unit displacement vectors $\vec{\delta}_{1}=a\hat{x}$ and $\hat{\delta}_{2}=a\hat{y}$ connect the atom $A$ with $B$ and $A$ with $C$, respectively. The corresponding nearest hopping parameters are $t_{1}$, $t_{2}$, whereas the next-nearest neighbor hopping parameter is $+it_{3}$ and $-it_{3}$ depending if it occurs in the direction denotes by the arrows.}
\label{fig:Lieb}
\end{figure}

Let us consider an electronic Lieb lattice~\footnote{The results here obtained apply to optical Lieb lattices as well.} so that the separation between two nearest atoms is $a$, the length of each side of the square is $\ell=2a$. There are three sites in the elementary cell, see Fig.~\ref{fig:Lieb-1}.  The primitive translation vectors are $\vec{r}_{1}=2a\hat{x}$ and $\vec{r}_{2}=2a\hat{y}$. It is customary to denote the atoms at the corners of the square as $A$, whereas the atoms at the sides of the square are $B$ (horizontal) and $C$ (vertical). The lattice vectors $\vec{\delta}_{1}=a \hat{x}=\vec{r}_{1}/2$ and $\vec{\delta}_{2}=a \hat{y}=\vec{r}_{2}/2$ connect  an atom on the site $A$ to those on the sites $B$ and $C$, respectively (see Fig.~\ref{fig:Lieb-2}). The atoms $A$, $B$ and $C$ form the three sublattices $\mathbf{R}_{A}=n_{1}\vec{r}_{1}+n_{2}\vec{r}_{2}$, $\mathbf{R}_{B}=\mathbf{\vec{R}}_{A}+\vec{\delta}_{1}$, and $\mathbf{R}_{C}=\vec{R}_{A}+\vec{\delta}_{2}$, respectively, with $n_{1},n_{2}\in\mathbb{Z}$. The reciprocal space is spanned by the translation vectors of the reciprocal space $\hat{r}_{k_{1}}$ and $\hat{r}_{k_{2}}$, $\hat{r}_{p}\cdot\hat{r}_{k_{q}}=2\pi\delta_{p,q}$, $p,q=1,2$. This leads to $\hat{r}_{k_{1}}=\frac{\pi}{a}\hat{x}$ and $\hat{r}_{k_{2}}=\frac{\pi}{a}\hat{y}$. The first Brillouin zone, constructed from the Wigner-Seitz rule, restricts to the region composed by $k_{x}\in[-\frac{\pi}{2a},\frac{\pi}{2a}]$ and $k_{y}\in[-\frac{\pi}{2a},\frac{\pi}{2a}]$.

The band structure of the electrons on the Lieb lattice can be analyzed with the use of the tight-binding model. There are considered the nearest neighbor (NN) interactions between the sites $A-B$ and $A-C$, represented by the hopping parameters $t_{1}$ and $t_{2}$, respectively. We take into account also the next-nearest neighbor (NNN) transition $B-C$, which can be complex valued, with the sign of phase dependent on the orientation of the hopping. This emerges due to external time-dependent driven fields in photonic Lieb lattices~\cite{Lon17}, and magnon Lieb and Kagome lattices~\cite{Owe18}. 

In particular, we consider a purely imaginary NNN hopping parameter $e^{\pm i\pi/2}t_{3}$, where the hopping phase is positive ($+$) is the hopping occurs counter-clock-wise, and negative ($-$) otherwise. Such a hopping dynamics is depicted in Fig.~\ref{fig:Lieb-1}. This type of hopping was introduced by Haldane in~\cite{Hal88} as a model for quantum anomalous Hall effect in graphene without strong external magnetic fields, which was latter found experimentally in~\cite{Cha13}. This configuration has been recently implemented in a dice lattice~\cite{Dey20} and experimentally in a honeycomb magnon lattice~\cite{Bos23} to explore the topological properties and phase transitions of such systems. See also~\cite{Xin23} for a recent review.
	
The spectral analysis of the tight-binding Hamiltonian reveals the triple-band spectrum \cite{Jak23},
\begin{equation}
	w_{0}(\vec{k})=0 , \quad w_{\pm}(\vec{k})=\pm2\sqrt{t_{1}^{2}\cos^{2}(ak_{x})+t_{2}^{2}\cos^{2}(ak_{y})+4t_{3}^{2}\sin^{2}(ak_{x})\sin^{2}(ak_{y})} .
\label{genericE}
\end{equation}

\begin{figure}
\centering
\subfloat[][$t_{3}=0$]{\includegraphics[width=0.3\textwidth]{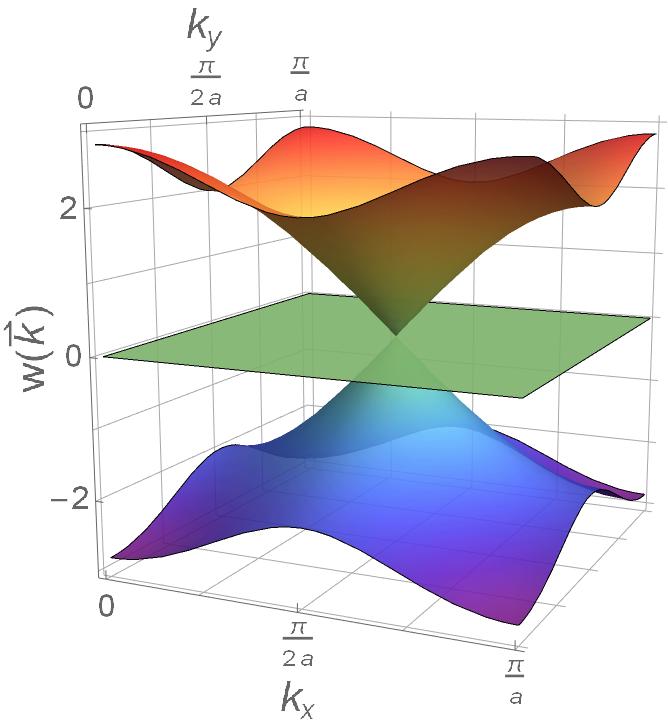}
\hspace{1mm}
\label{fig:DP0}}
\subfloat[][$t_{3}\neq 0$]{\includegraphics[width=0.3\textwidth]{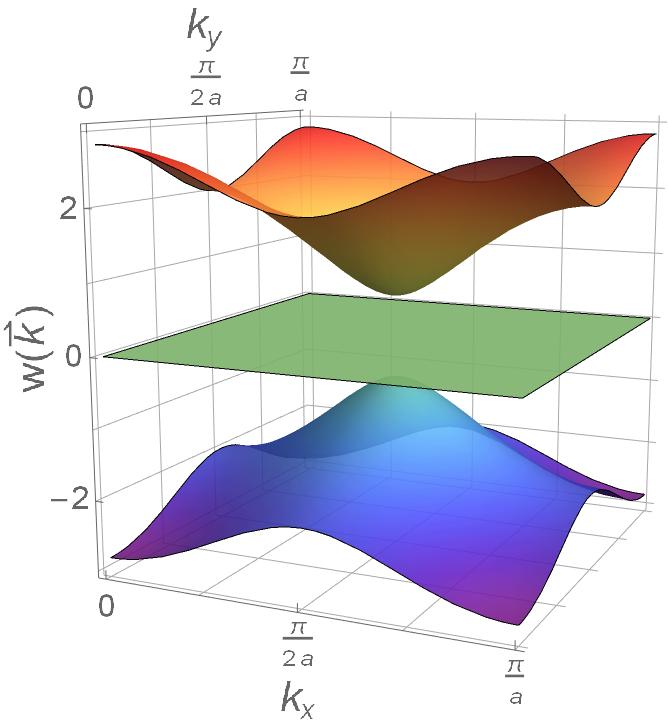}
\label{fig:DP1}}
\caption{Dispersion bands $w_{+}(\vec{k})$ (yellow-upper), $w_{-}(\vec{k})$ (green-lower), and $w_{0}(\vec{k})$ (blue-middle) for the gapless (a) and gapped (b) configurations.}
\label{fig:Bands}
\end{figure}

The bands have linear dependence on the momentum around the Dirac point situated in the first Brillouin zone, the explicit position of which depends on the relative strength of $t_3$. In this work, we focus on the most relevant situation where $t_{3}<\frac{t_{1}}{2}, t_{3}<\frac{t_{2}}{2}$. In that case, the Dirac point is $\vec{K}=(\frac{\pi}{2a},\frac{\pi}{2a})$, see Fig.\ref{fig:DP1} for illustration. A similar analysis holds for higher values of $t_3$, where the Dirac points are displaced with respect to $\vec{K}$. For a detailed discussion, see \cite{Jak23}.  

Let us calculate the approximate form of the tight-binding Hamiltonian in the vicinity of the Dirac point $\vec{K}$. We denote the effective operator as $\mathcal{H}(\vec{k})\equiv \mathcal{H}(\vec{K}+\vec{k})$, where $\vert\vec{k}\vert$ is considered small enough so that we can keep terms up to first-order in $\vec{k}$. The proper expansion of $\mathcal{H}(\vec{k})$ at the Dirac point $\vec{K}$ can be conveniently written as
\begin{equation}
\mathcal{H}(\vec{k})= 2a t_{1} k_{x} S_{1} + 2a t_{2} k_{y} S_{2} + 4 t_{3} S_{3} .
\label{Dirac-H1}
\end{equation}
The matrices 
\begin{equation}
	S_{1}=
	\begin{pmatrix}
	0 & 1 & 0 \\
	1 & 0 & 0 \\
	0 & 0 & 0
	\end{pmatrix}
	, \quad 
	S_{2}=
	\begin{pmatrix}
	0 & 0 & 1 \\
	0 & 0 & 0 \\
	1 & 0 & 0
	\end{pmatrix} 
	, \quad 
	S_{3}=
	\begin{pmatrix}
	0 & 0 & 0 \\
	0 & 0 & -i \\
	0 & i & 0
	\end{pmatrix}
	,
	\label{matrix-spin1}
	\end{equation}
form the three-dimensional representation of $su(2)$ algebra, $[S_{p},S_{q}]=i\varepsilon_{pqr}S_{r}$, with $\varepsilon_{pqr}$ the three-dimensional anti-symmetric tensor. Therefore, the quasi-particles described by the effective Hamiltonian (\ref{Dirac-H1}) have pseudospin 1.
	
It is worth noting that, for $t_{3}=0$, the resulting Dirac Hamiltonians in~\eqref{Dirac-H1} becomes linear combinations of the spin-1 matrices $S_{1}$ and $S_{2}$. In such a case, the matrix $\widetilde{S}$,
	\begin{equation}\label{widetildeS}
	\widetilde{S}=\begin{pmatrix}
	-1 & 0 & 0 \\
	0 & 1 & 0 \\
	0 & 0 & 1 
	\end{pmatrix},
	\end{equation}
	satisfies $\{\widetilde{S},S_{j}\}=0$, with $j=1,2$, and represents the chiral symmetry of $\mathcal{H}$ as there holds $\{\widetilde{S},\mathcal{H}\vert_{t_{3}=0}\}=0$. The later relation implies that the eigenvalues $E$ of $\mathcal{H}\vert_{t_{3}=0}$ are symmetric with respect to $E=0$. When an eigenstate $\boldsymbol{\Psi}_{E}$ of $\mathcal{H}$ has energy $E$, then there is an eigenstate $\boldsymbol{\Psi}_{-E}=\widetilde{S}\boldsymbol{\Psi}_{E}$ with the energy of the opposite sign. 
	
\subsection{External electrostatic interaction}
\label{subsec:Spin1D}
Throughout this manuscript, we consider a piece-wise continuous external electric field distributed in the $\hat{x}$ direction, while we discard any magnetic interaction. The corresponding effective Hamiltonian is obtained from \eqref{Dirac-H1} through the Peierls transformation~\cite{Pei29,Blo28}, $\vec{k}\rightarrow-i\hbar\vec{\nabla}$ and $i\hbar\partial_{t}\rightarrow i\hbar\partial_{t}-U(\vec{x})\mathbb{I}$, with $\mathbb{I}$ the $3\times 3$ identity matrix. Since the Hamiltonian becomes invariant on the $\hat{y}$ direction, the eigenstates can be cast in the form $\boldsymbol{\Psi}(x,y)\rightarrow e^{\pm i k_{2}y}\boldsymbol{\Psi}(x)$, where $\boldsymbol{\Psi}(x)$ solve the following stationary equation:
\begin{equation}
H(x)\boldsymbol{\Psi}(x)= (-i\hbar v_{1}S_{1}\partial_{x} + \hbar v_{2} k_{y} S_{2} + m S_{3} +U_{a}\,\mathbb{I})\boldsymbol{\Psi}(x)=E\boldsymbol{\Psi}(x) ,
\label{Dirac-H2}
\end{equation}
with $\boldsymbol{\Psi}(x)=(\psi_{A}(x),\psi_{B}(x),\psi_{C}(x))^{T}$.

In~\eqref{Dirac-H2}, we have used $v_{1}=2at_{1}$, $v_{2}=2at_{2}$ and $m=4t_{3}$ to simplify the notation. This allows us relating $v_{1}$ and $v_{2}$ to the Fermi velocities along the $\hat{x}$ and $\hat{y}$ directions, respectively, whereas $m$ plays the role of the mass term in the Dirac equation. Furthermore, we have considered a constant electrostatic potential, which is valid for our purposes since we are dealing with piece-wise continuous interactions.

From the previous considerations, we may decouple the eigensolution components $\psi_{A,B,C}$ as follows:
\begin{align}
&-\hbar^{2}v_{1}^{2}\psi_{A}'' + \hbar^{2}v_{2}^{2}k_{y}^{2}\psi_{A}=((E-U_{a})^{2}-m^{2})\psi_{A} , 
\label{psiA}\\
& \psi_{B}=-i\frac{\hbar v_{1}(E-U_{a})\psi_{A}'+\hbar m v_{2} k_{y}\psi_{A}}{(E-U_{a})^{2}-m^{2}} , \quad \psi_{C}=\frac{\hbar m v_{1} \psi_{A}'+\hbar v_{2}k_{y}(E-U_{a})\psi_{A}}{(E-U_{a})^{2}-m^{2}} ,
\label{psiB}
\end{align}
where the hopping parameters ${t}_{j}$, for $j=1,2,3$.
	
The probability current associated with $\boldsymbol{\Psi}$ can be calculated in standard manner from the continuity equation $\partial_{t}\rho+\vec{\nabla}\cdot\boldsymbol{j}=0$. Here, $\rho=\boldsymbol{\Psi}^{\dagger}\boldsymbol{\Psi}$ stands for the probability density, and the probability current takes the form
\begin{equation}
\boldsymbol{j}=\left(2v_{1}\operatorname{Re}\psi_{A}^{*}\psi_{B},2v_{2}\operatorname{Re}\psi_{A}^{*}\psi_{C} \right) .
\label{current}
\end{equation}
	
Let us consider briefly the situation when the potential has a finite discontinuity at $x=x_0$. It is necessary to specify the behavior of the wave functions at this point. It can be done by integrating (\ref{Dirac-H2}) around the vicinity of $x_0$. Alternatively, one can require the component of the density current perpendicular to the barrier to be continuous. The second approach is more general and covers the boundary conditions provided by the integration as the special case that read as
\begin{equation}
\psi_{A}(x_{0}^{-})=\psi_{A}(x_{0}^{+}), \quad \psi_{B}(x_{0}^{-})=\psi_{B}(x_{0}^{+}). 
\label{cont}
\end{equation}
It is worth noting that only two of the three eigensolution components are required to be continuous in $x_0$, and the third component $\psi_{C}(x)$ can have a discontinuity at this point. The corresponding probability density is not necessarily continuous, an observation already made for pseudospin-1 photonic lattices~\cite{Xu17}. The boundary conditions obtained in~\eqref{cont} keep the probability current in the $\hat{x}$ direction continuous, which in our case corresponds to the component perpendicular to the discontinuity. Although the component $\Psi_C(x)$ can be discontinuous at $x_0$, one finds from the third component of~\eqref{Dirac-H2} that $(E-U(x))\psi_{C}(x)$ shall be continuous at $x=x_{0}$. Despite the latter, the current tangent to the discontinuity is not necessarily continuous.

\section{Rectangular electrostatic barrier}
\label{sec:null-mag}
Let us consider an external electrostatic electric potential homogeneous along the $\hat{y}$ direction and piece-wise continuous across the $\hat{x}$ direction, with
\begin{equation}
U(x)=\begin{cases}
0 \quad &\vert x\vert>\frac{L}{2} \\
U_{0} \quad &\vert x\vert\leq \frac{L}{2}
\end{cases}
.
\label{U-pot}
\end{equation}
We consider, without loss of generality, $U_{0}>2m$. Solutions of the stationary equation are split into three regions, namely, the \textit{region} I ($x<L/2$), \textit{region} II ($-L/2 \leq x \leq L/2$), and \textit{region} III ($x>L/2$). The latter are written as
\begin{equation}
\boldsymbol{\Xi}_{k_a}=e^{i k_{y}y}e^{ik_a x}
\begin{pmatrix}
1 \\
\frac{-i \hbar m v_2 k_{y}+\hbar v_1k_a(E-U_a)}{(E-U_a)^2-m^2}\\
\frac{i \hbar m v_{1} k_a +\hbar v_2k_{y}(E-U_a)}{(E-U_a)^2-m^2}
\end{pmatrix}
,\quad a=\operatorname{I},\operatorname{II},\operatorname{III},
\label{fp-gen}
\end{equation}
where $U_{\operatorname{I}}=U_{\operatorname{III}}=0,$ $U_{\operatorname{II}}=U_0$, and consequently $k_{\operatorname{I}}=k_{\operatorname{III}}$. We consider $k_y$ is fixed as a real quantity to obtain plane-wave solutions propagating parallel to the barrier. In turn, $k_a$ is considered a complex parameter so that we can distinguish two different regimes (see discussion below). 

The solution (\ref{fp-gen}) satisfies the eigenvalue equation~\eqref{Dirac-H2} with the eigenvalue
\begin{equation}
E=U_{a}\pm\sqrt{m^{2}+\hbar^{2}(v_{1}^{2}k_a^{2}+v_{2}^{2}k_{y}^{2})}=U_{a}\pm\sqrt{\tilde{m}^{2}+\hbar^{2}v_{1}^{2}k_a^{2}},
\label{E}
\end{equation}
where we have introduced the \textit{effective mass} term
\begin{equation}
\widetilde{m}=\sqrt{m^{2}+\hbar^{2}v_{2}^2k_{y}^{2}} .
\label{mt}
\end{equation} 

\begin{figure}
\centering
\subfloat[][]{\includegraphics[width=0.33\textwidth]{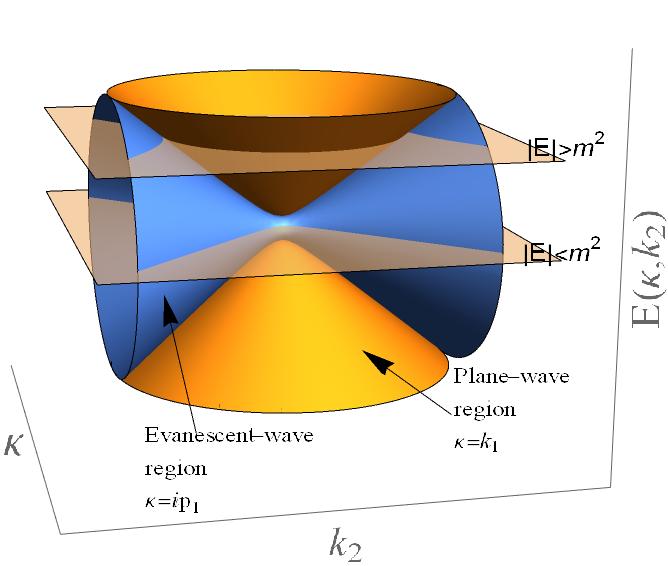}
\label{fig:surface-0}}
\hspace{2mm}
\subfloat[][$\vert E\vert>m$]{\includegraphics[width=0.27\textwidth]{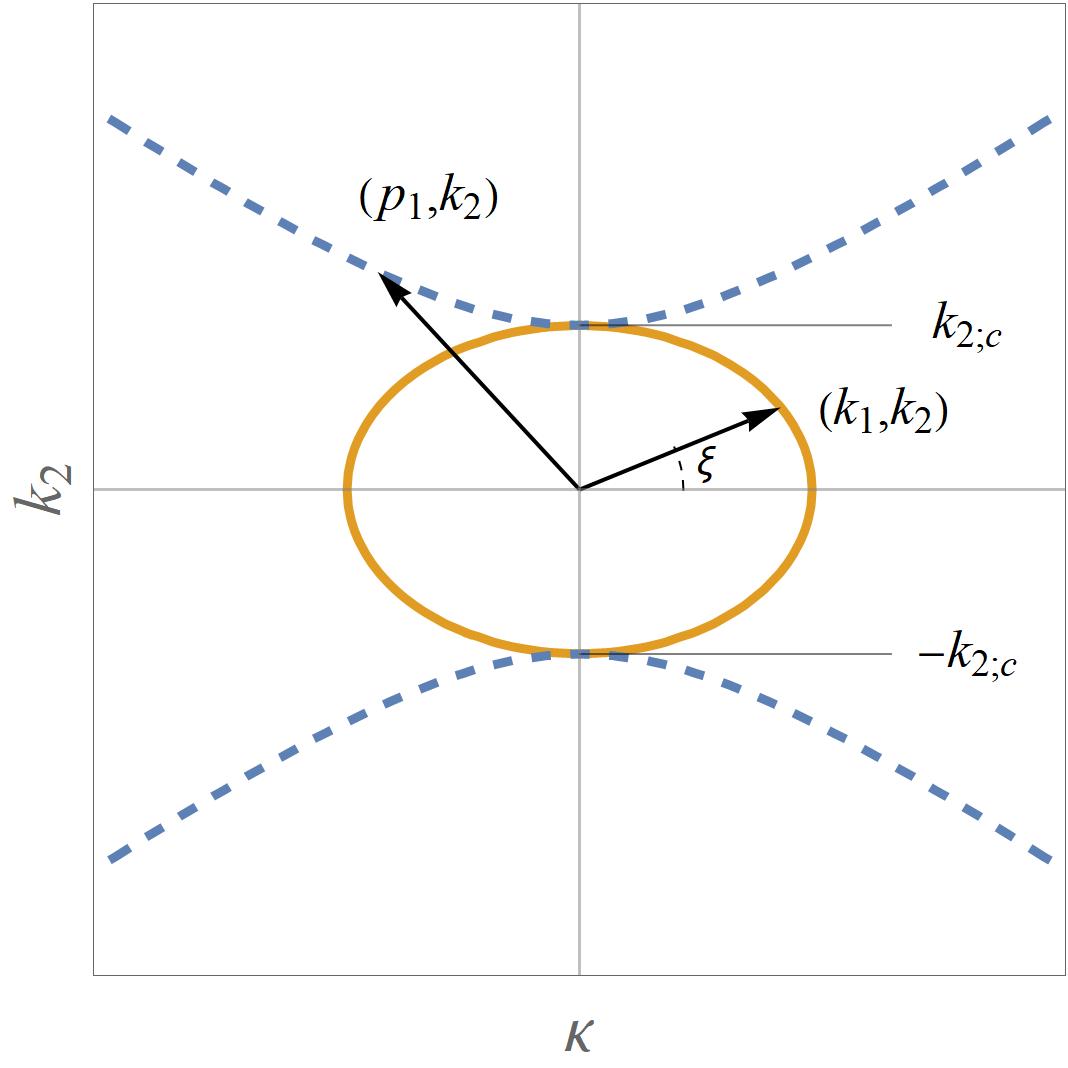}
\label{fig:surface-1}}
\hspace{2mm}
\subfloat[][$\vert E\vert<m$]{\includegraphics[width=0.27\textwidth]{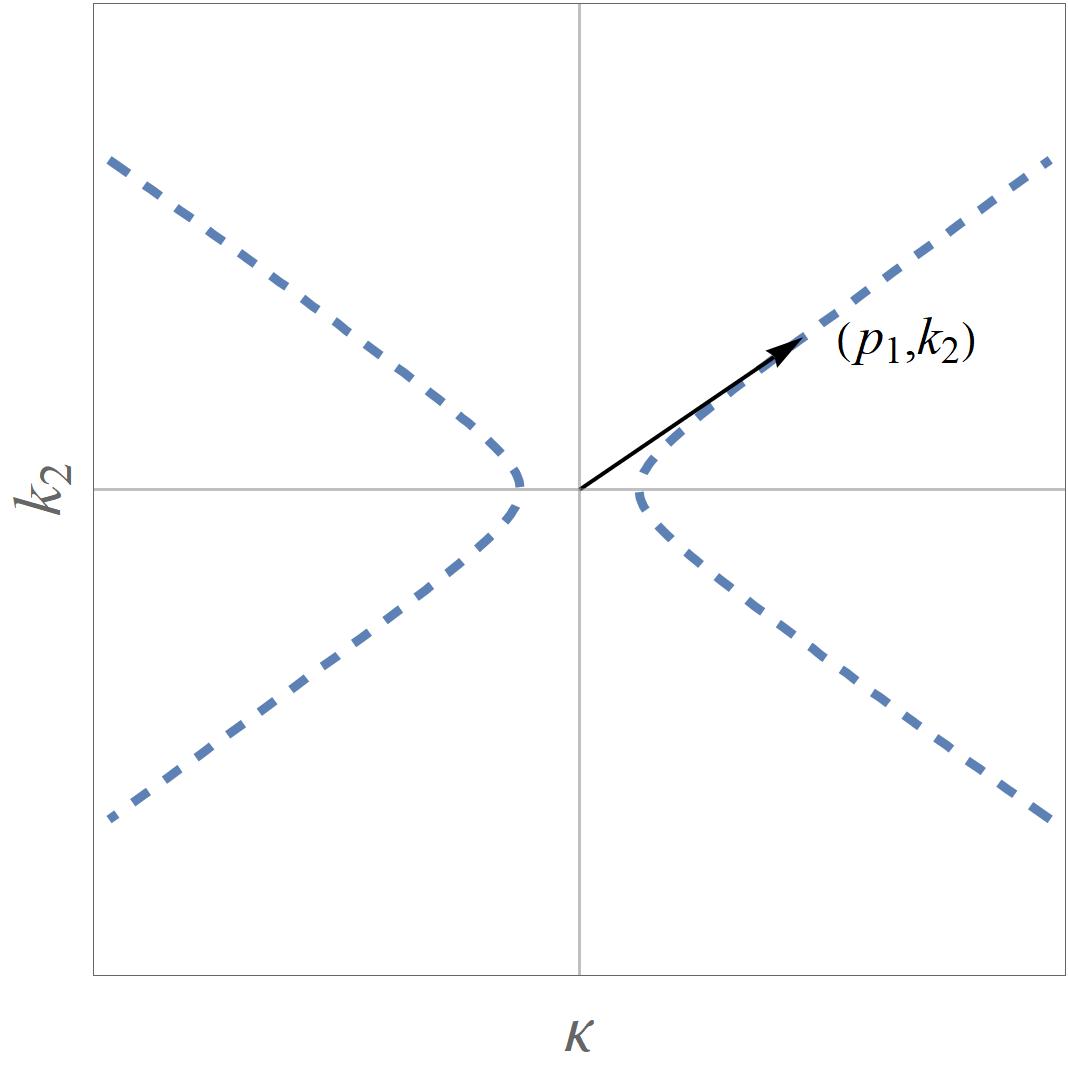}
\label{fig:surface-2}}
\caption{(a) Sketch of the energy surfaces spanned by the dispersion relations~\eqref{en1} (orange) and~\eqref{en2} (blue), together with two energy planes located at arbitrary energies $\vert E\vert>m$ and $\vert E\vert<m$. Panel (b) and panel (c) depict the contour plot generated by the interception between the dispersion relations and the energy planes $\vert E\vert>m$ and $\vert E\vert<m$, respectively. In panel (b), $\xi=\arctan(k_{y}/k_{\operatorname{I}})$ denotes the incidence angle of the plane wave and $k_{y;c}=\sqrt{E^{2}-m^{2}}/\hbar v_{2}$ the critical value of $k_{y}$ separating the evanescent-wave and plane-wave regimes.}
\label{fig:Energy-surface}
\end{figure}

From~\eqref{fp-gen}, we distinguish two behaviors, namely, plane-wave solutions for $k_a\in\mathbb{R}$ and evanescent-wave solutions for $k_a=-ip_{a}$, with $p_a\in\mathbb{R}$. In both cases, the wave functions are associated with real eigenvalues. They are classified as
\begin{align}
&k_a\in \mathbb{R}, \quad && E(k_{a},k_{y})=U_{a}\pm\sqrt{\widetilde{m}+v_{1}^{2}\hbar^{2}k_{a}^{2}}, && E(p_{a},k_{y})\in(-\infty,U_{0}-\widetilde{m})\cup(U_{a}+\widetilde{m},\infty),  
\label{en1}\\
& k_a=-i\, p_{a}, \quad && E(p_{a},k_{y})=U_{a}\pm\sqrt{\widetilde{m}-v_{1}^{2}\hbar^{2}p_{a}^{2}}, && E(p_{a},k_{y})\in(U_{a}-\widetilde{m},U_{0}+\widetilde{m}).
\label{en2}
\end{align}
These dispersion relations span paraboloid and hyperboloid surfaces for plane-wave and evanescent-wave solutions, respectively. This behavior is depicted in Fig.~\ref{fig:surface-0} for $U_{\operatorname{I}}=0$ (case $a=\operatorname{I}$).

For $\vert E\vert>m$, the behavior of the solutions is classified according to the values of $k_{y}$, with $k_{y;c}=\sqrt{E^{2}-m^{2}}/\hbar v_{2}$ being the critical value. That is, for $\vert k_{y}\vert<k_{y;c}$, the solutions are plane-wave-like and the momenta $k_{\operatorname{I}}$ and $k_{y}$ span an elliptic curve for a fixed energy. For $\vert k_{y}\vert>k_{y;c}$,  the solutions become evanescent waves and $p_{\operatorname{I}}$ and $k_{y}$ span a hyperbolic curve for the same fixed energy. This is sketched in Fig.~\ref{fig:surface-1}. For $\vert E\vert<m$, no plane-wave solutions exist for $k_{y}\in\mathbb{R}$, and only evanescent-wave solutions are generated. Here, $p_{\operatorname{I}},k_{y}$ span a rotated hyperbola with respect to the case $\vert E\vert>m$, as depicted in Fig.~\ref{fig:surface-2}. 

The partial solutions $\Xi_{k_a}$ at the regions $\operatorname{I},\ \operatorname{II},\ \operatorname{III}$ have to be combined in order to comply with the boundary conditions (\ref{cont}) at $x_0=\pm L$ and $\vert x\vert\rightarrow\infty$. 
The wave function takes the general form
\begin{equation}
\Xi_a(x,y)=\alpha_a\Xi^{\pm}_{k_a}(x,y)+\beta_a\Xi^{\pm}_{-k_a}(x,y),\quad a=\operatorname{I},\ \operatorname{II},\ \operatorname{III}.\label{Xi}
\end{equation}
The boundary conditions (\ref{cont}) impose the continuity of the two upper components of the wave function, from which we find the set of relations between the coefficients $\alpha_{\operatorname{I}}$, $\beta_{\operatorname{I}}$ and $\alpha_{\operatorname{III}}$, $\beta_{\operatorname{III}}$. That is,
\begin{equation}\label{transfermatrix}
\mathbb{M}\begin{pmatrix}\alpha_{\operatorname{I}}\\ \beta_{\operatorname{I}}\end{pmatrix}=\begin{pmatrix}\alpha_{\operatorname{III}}\\ \beta_{\operatorname{III}}\end{pmatrix}, \quad \mathbb{M}=\begin{pmatrix}m_{11}&m_{12}\\m_{21}&m_{22}\end{pmatrix} ,
\end{equation}
with $\mathbb{M}$ being the \textit{transfer matrix}, whose elements $m_{ij}$ are functions of $E$, $k_{y}$, $m$ and $U_0$. The latter are explicitly given by
\begin{align}
&m_{11}=e^{i L k_{\operatorname{I}}}\left(\cos( L k_{\operatorname{II}})-i \sin (Lk_{\operatorname{II}})
\frac{2v_1^2\hbar^2 k_{\operatorname{I}}^2k_{\operatorname{II}}^2+\tilde{m}^2(k_{\operatorname{I}}^2+k_{\operatorname{II}}^2)}{2E(E-U_{0})k_{\operatorname{I}}k_{\operatorname{II}}}\right),\\
&m_{22}=e^{-i L k_{\operatorname{I}}}\left(\cos (L k_{\operatorname{II}})+i \sin (Lk_{\operatorname{II}})
\frac{2v_1^2\hbar^2 k_{\operatorname{I}}^2k_{\operatorname{II}}^2+\tilde{m}^2(k_{\operatorname{I}}^2+k_{\operatorname{II}}^2)}{2E(E-U_{0})k_{\operatorname{I}}k_{\operatorname{II}}}\right),\\
&m_{12}=i\frac{\sin(L k_{\operatorname{II}})(k_{y}^2v_2^2(m^2+\hbar^2k_{y}^2v_2^2)-v_1^2(m^2-\hbar^2 k_y^2 v_2^2)k_1^2-2iv_1 v_2 k_{\operatorname{I}}k_{y} E)(\frac{U_0}{2}-E)}{\hbar^2v_1^2k_{\operatorname{I}}k_{\operatorname{II}}(E^2-m^2)(E-U_0)},
\end{align}
where $k_{\operatorname{I}}=\sqrt{E^2-\tilde{m}^2}/\hbar v_{1}$ and $k_{\operatorname{II}}=\sqrt{(E-U_{0})^2-\tilde{m}^2}/\hbar v_{1}$. 

The determinant of the transfer matrix is equal to one, in coherence with conservation of the probability current at the boundary. When $k_{\operatorname{I}}$ and $k_{\operatorname{II}}$ are real, there also holds $m_{11}=m_{22}^*$ and $m_{12}=m_{21}^*$. In the next section, we shall use the transfer matrix for determinantion of the bound state energies as well as of the scattering characteristics of the plane-wave solutions. 

Additionally to (\ref{fp-gen}), the Lieb lattice supports an additional solution in the form of a flat band, which is depicted in Fig.~\ref{fig:Bands}. This appears whenever $E=U_{a}$, and the eigensolutions cannot be determined from Eq.~\eqref{psiA}-\eqref{psiB}. Instead, one shall solve the Dirac Hamiltonian for $E=U_{a}$, which leads to the eigensolution $\boldsymbol{\Xi}_{fb}=(m\chi,i\hbar v_{2}k_{y}\chi,-\hbar v_{1} \chi')^{T}$, where $\chi=\chi(x)$ is an arbitrary complex-valued function. Such an indeterminacy is better understood if one chooses $\chi$ such that
\begin{equation}
\boldsymbol{\Xi}_{fb}(\nu,x)=e^{ik_{y}y}e^{i\nu x}
\begin{pmatrix}
m \\
i\hbar v_{2}k_{y} \\
-i\hbar v_{1} \nu
\end{pmatrix}
,
\label{fb-sol}
\end{equation}
which is a flat band eigensolution for any $\nu\in\mathbb{C}$. Particularly, for $\nu\in\mathbb{R}$, Eq.~\eqref{fb-sol} form a set of degenerate plane-wave solutions, usually known as \textit{degenerate Bloch waves}~\cite{Aok96} (see also Sec. 2.1 in~\cite{Ley18}). These degenerate waves form a continuous basis that can be used to construct arbitrary wavepackets through Fourier transforms. The latter has been exploited to construct the so-called \textit{compact localizes states}~\cite{Ber08}, which are specific linear combinations of degenerate waves localized in each unitary cell of a finite-dimensional lattice. See~\cite{Ley18,Rhi21} for a more extensive discussion on the matter.

It is clear that degenerate Bloch waves do not carry current on the $x$-direction, as $\boldsymbol{j}_{x}=2v_{1}\operatorname{Re}\psi_{A}^{*}\psi_{B}$ vanishes for any $\nu\in\mathbb{C}$. This also holds for any linear combination (finite or infinite) of degenerate Bloch state. Thus, the current states belonging to the flat band energy are current-free states. 

Notice that the dispersion and flat bands have a touching point only for $m=0$ (See Fig.~\ref{fig:DP0}), and thus one can explore the behavior of the solutions on the dispersion band when they approach the flat band interception. It is straightforward to realize that $\Xi_{k_a,k_{y},m\rightarrow 0}$ leads to the null vector, which is only one of the infinitely many solutions inside the flat band. For this reason, we shall discuss the flat band and the dispersion bands separately.

\section{Electron confinement}
\label{subsec:U-confinement}
Let us explore the possibility of bound states trapped by the electrostatic potential~\eqref{U-pot}. Here, we look for eigenvalues $E$ so that the corresponding eigensolutions have finite norm in $L^{2}\otimes\mathbb{C}^{3}$, which implies that eigensolutions must decay asymptotically to zero in the regions I and III for $x\rightarrow-\infty$ and $x\rightarrow\infty$, respectively. 

\begin{figure}
	\centering
	\subfloat[][$k_{y}=0$]{\includegraphics[width=0.3\textwidth]{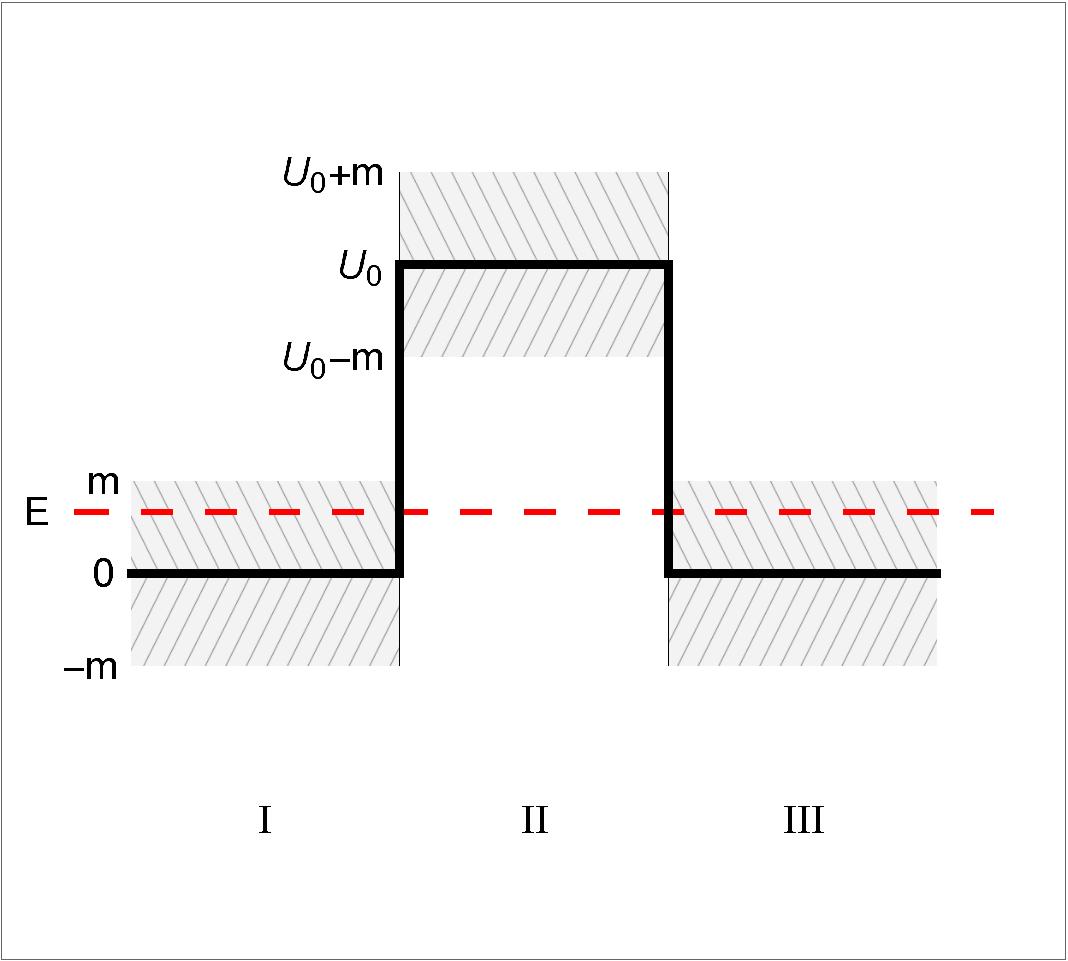}
		\label{fig:step0}}
	\hspace{2mm}
	\subfloat[][]{\includegraphics[width=0.3\textwidth]{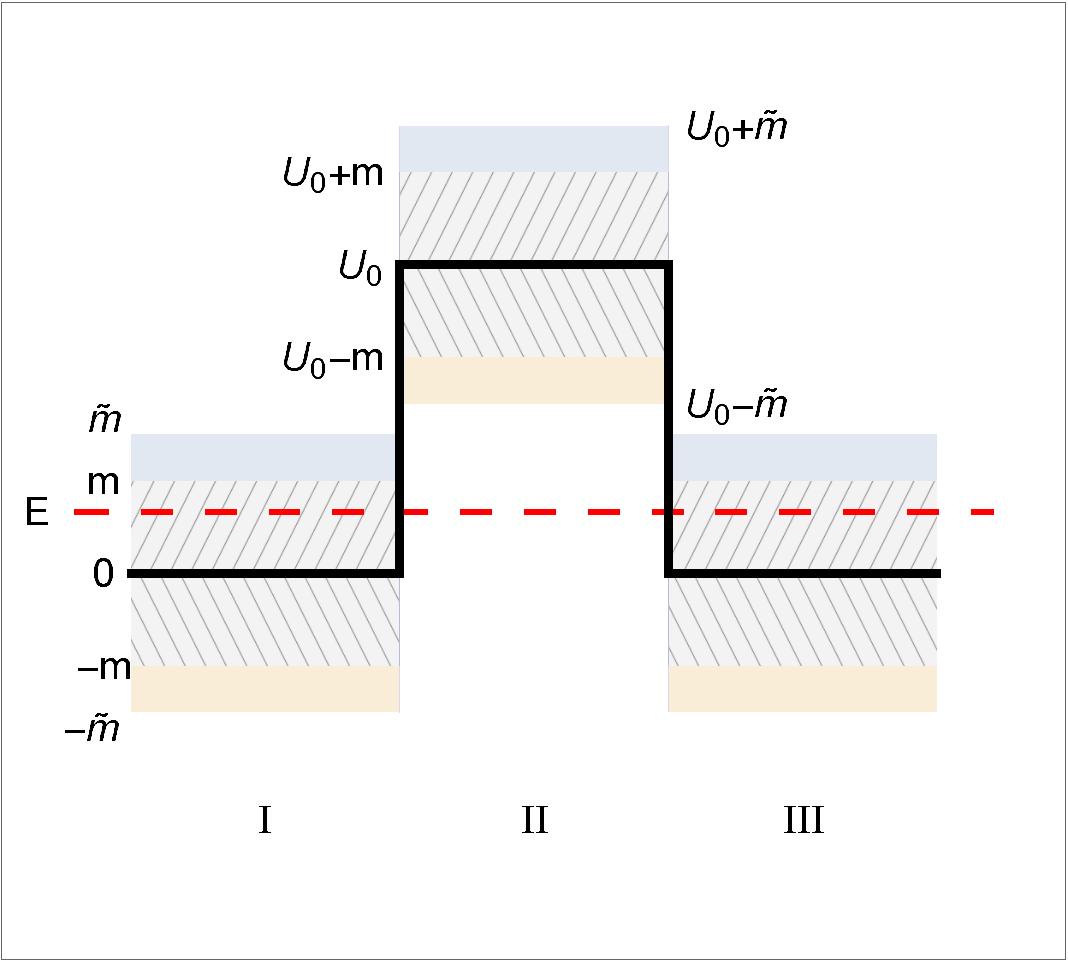}
		\label{fig:step1}}
	\hspace{2mm}
	\subfloat[][]{\includegraphics[width=0.3\textwidth]{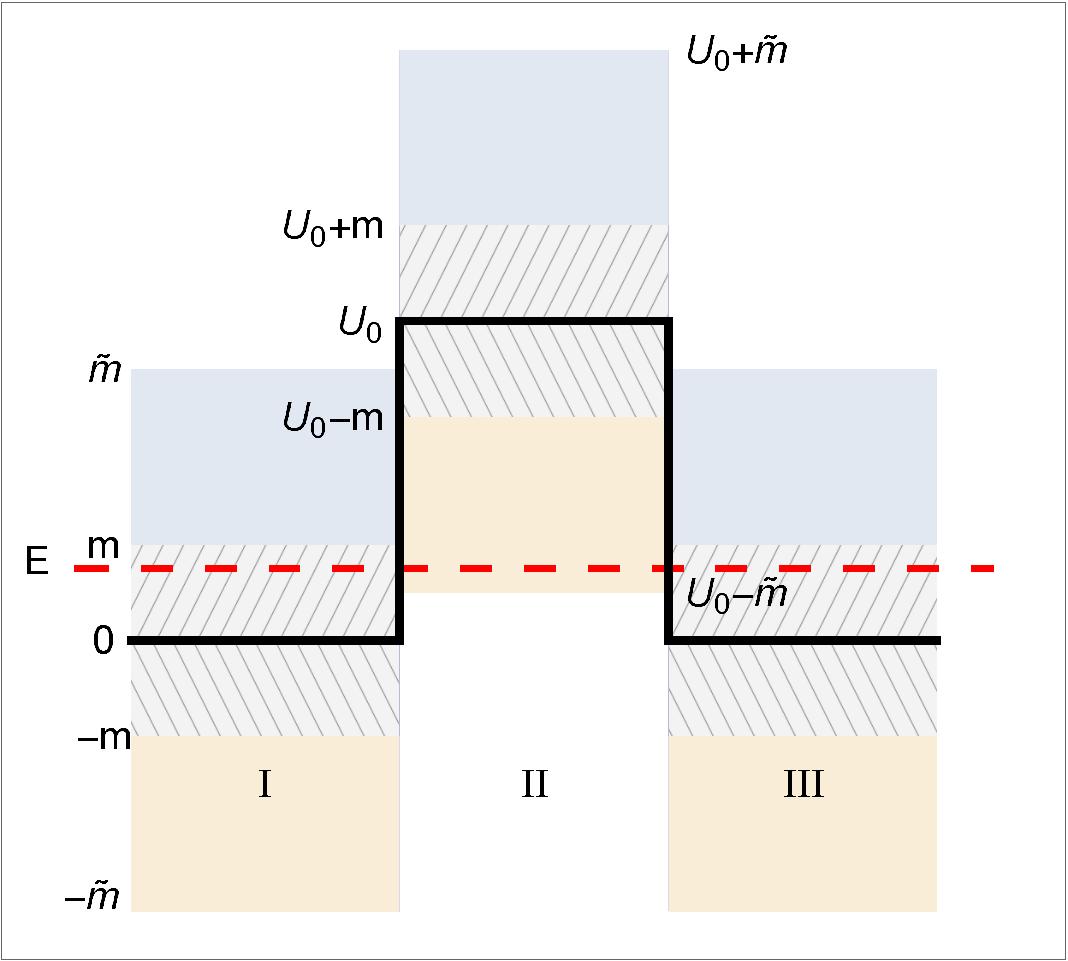}
		\label{fig:step2}}
	\caption{Sketch for the energy configuration associated with~\eqref{U-pot} for $k_{y}=0$ (a) and increasing values of $k_{y}$ (b)-(c). The diagonal-pattern and color-shaded regions denote the area covered by mass term $m$ and effective mass term $\widetilde{m}$, respectively. In the panel (b), the energy (red-dashed line) inside the region II lies out of the effective mass term (plane-wave solution), whereas in the panel (c) they lie inside the effective mass term (evanescent-wave solution).}
	\label{fig:bound-diagram}
\end{figure}

Following~\eqref{fp-gen}, we thus use evanescent-wave solutions for the regions I and III. By fixing
\begin{equation}
k_{\operatorname{I}}=k_{\operatorname{III}}=ip_{\operatorname{I}},\quad p_{\operatorname{I}}>0,
\end{equation}
one restricts the energies into the interval $E\in(-\widetilde{m},\widetilde{m})$, as depicted in all the cases of Fig.~\ref{fig:bound-diagram}. 
The wave function composed from (\ref{Xi}) has an exponentially vanishing behavior for $|x|\rightarrow\infty$. This implies that we fix $\alpha_{\operatorname{I}}=0$, $\beta_{\operatorname{I}}=1$ and $\beta_{\operatorname{III}}=0$, and the relation (\ref{transfermatrix}) turns into
\begin{equation}
m_{12}=\alpha_{\operatorname{III}},\quad m_{22}=0.
\end{equation}
The first relation determines the amplitude of the wave function in the region III, whereas the second relation fixes the energies for the bound states. This can be written, after some simplifications, in the following form:
\begin{equation}
\tanh\left(\frac{\sqrt{\widetilde{m}^{2}-(E-U_{0})^{2}}}{\hbar v_{1}} L \right)=-\frac{E(E-U_{0})\sqrt{\widetilde{m}^{2}-(E-U_{0})^2}\sqrt{\widetilde{m}^{2}-E^{2}}}{(E-U_{0})^{2}(\widetilde{m}^{2}-E^{2})+\widetilde{m}^{2}U_{0}\left(E-\frac{U_{0}}{2}\right)}.
\label{trans}
\end{equation}

The wave function $\Xi_{\operatorname{II}}$ in the intermediate region II can be either oscillatory for $(E-U_0)^2>\tilde{m}^2$ (we can set $k_{\operatorname{II}}>0$ without loss of generality) or evanescent for $(E-U_0)^2<\tilde{m}^2$ ($k_{\operatorname{II}}=i\,p_{\operatorname{II}}$, $p_{\operatorname{II}}>0$), see Fig.~\ref{fig:step1} and Fig.~\ref{fig:step2}, respectively. The transcendental equation~\eqref{trans} allows us determining the bound state energies as a function of $k_{y}$ for both cases.

Although the explicit solution $E=E(k_{y})$ of (\ref{trans}) has to be found numerically, some preliminary information can be extracted by considering large values $\hbar v_{2}k_{y}\gg U_{0},m$ in the transcendental equation~\eqref{trans}. Here, $\widetilde{m}\approx \hbar v_{2}k_{y}$ and the dispersion relation reduces to $E^{2}\approx E^{2}_{\infty} = \hbar^{2}(-v_{1}^{2}p_{\operatorname{I}}^{2}+v_{2}^{2}k_{y}^{2})$. Since $p_{\operatorname{I}}$ should be a real quantity in order to remain in the evanescent-wave regime in the regions I and III, we find that $\hbar v_{2}\vert k_{y}\vert\geq E(k_{y})$ holds for asymptotic values of $\hbar v_{2} k_{y}$. The behavior of $E(k_{y})$ is thus bounded for $\hbar v_{2}k_{y}\rightarrow\infty$ and can be classified into the following in three asymptotic cases:
\begin{itemize}

\item First, a valid asymptotic behavior may be of the form $E(k_{y}\rightarrow\infty)\rightarrow C<\infty$. Substituting the latter into~\eqref{trans} leads to a unique solution of the form $E(k_{y}\rightarrow\infty)\rightarrow C=U_{0}/2$.

\item Another possible asymptotic behavior is $\vert E(k_{y})\vert =\hbar v_{2} \vert k_{y}\vert$, which vanishes both sides of~\eqref{trans}. That is, $\vert E(k_{y})\vert =\hbar v_{2} \vert k_{y}\vert$ is a valid asymptotic behavior.

\item The last possible asymptotic behavior is $\vert E(k_{y})\vert < \hbar v_{2} \vert k_{y}\vert$, which leads to a contradiction once substituted into~\eqref{trans}. That is, such an asymptotic behavior is do not generate bound state solutions.

\end{itemize}

We thus conclude that the eigenvalues associated with bound states, if they exist, either converge asymptotically to $U_{0}/2$ or $\hbar v_{2} \vert k_{y}\vert$. Since the current density on the direction parallel to the barrier is $J_{y}=\partial E(k_{y})/\partial k_{y}\equiv\int_{\mathbb{R}}\boldsymbol{j}(x,y)dx$ (see \cite{Gho08} or Appendix E in~\cite{Ash76}), it converges either to zero or $\pm \hbar v_{2}$ for $k_{y}\rightarrow\infty$. 

\begin{figure}
	\centering
	\subfloat[][]{\includegraphics[width=0.3\textwidth]{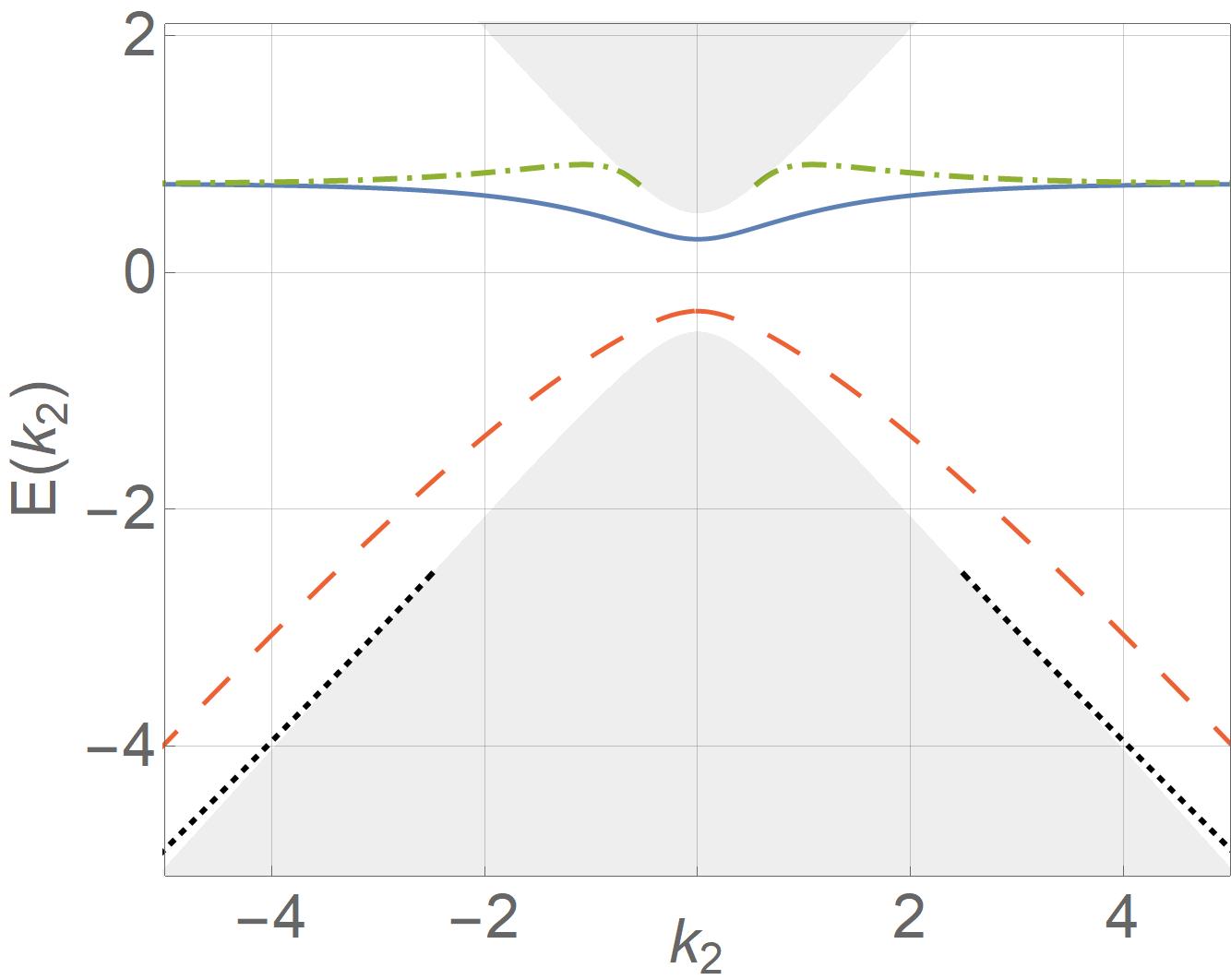}
		\label{fig:bound-k2-E}}
	\hspace{5mm}
	\subfloat[][]{\includegraphics[width=0.3\textwidth]{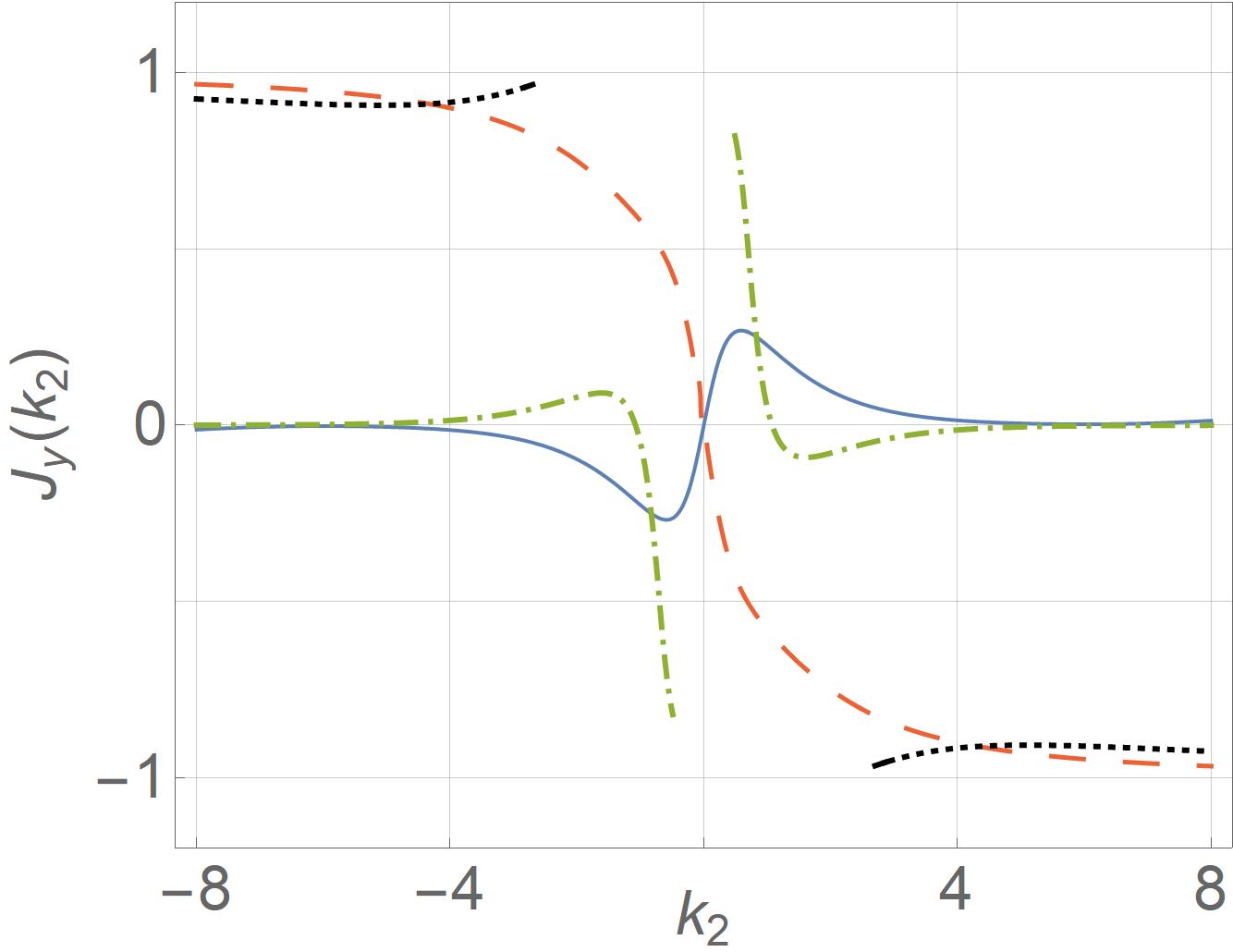}
		\label{fig:bound-k2-current}}
\caption{(In units of $\hbar$=1) (a) Bound state energies $E(k_{y})$, computed from~\eqref{trans}, as a function of the transverse momentum $k_{y}$ for $v_{1}=v_{2}=L=1$, $m=0.5$, and $U_{0}=1.5$. The blue-solid and red-dashed curves indicate bound state energies for arbitrary $k_{y}$, whereas green-dot-dashed and black-dotted curves are energies emerging from a specific $k_{y}\neq 0$. The shaded area marks the scattering-state energy region. (b) Current parallel to the barrier $J_{y}=\partial E(k_{y})/\partial k_{y}$ associated with the dispersion relations in (a).}
	\label{fig:bound-k2}
\end{figure}

As an illustrative example, let us consider numerical values such that we have homogeneous Fermi velocities $v_{1}=v_{2}=1$, a mass term $m=0.5$, together with a rectangular potential well with $L/\hbar=1$ and $U_{0}=1.5$. Numerical solutions of~\eqref{trans} reveal the existence of two bound states for $k_{y}=0$\footnote{This result agrees with the analytic formula presented in~\eqref{N-even-odd}.}, and new bound states appear for increasing values of $k_{y}$. This is depicted in Fig.~\ref{fig:bound-k2-E}, where one may see that energies indeed converge to either $\tfrac{U_{0}}{2}=0.75$ or become linear in $\hbar v_{2} k_{y}$ for large enough $k_{y}$. Likewise, we depict in Fig.~\ref{fig:bound-k2-current} the corresponding current density parallel to the barrier ($J_{y}$), which becomes finite or null for asymptotic $k_{y}$, as predicted from our former analysis.

$\bullet$ Further information is available for direct incidence, that is, $k_{y}=0, \widetilde{m}= m$. Here, the effective Hamiltonian possesses the additional symmetry represented $[\mathcal{H},P_x\widetilde{S}]=0$, with $P_x$ is the parity operator and $\widetilde{S}$ defined in~\eqref{widetildeS}. This allows establishing a parity-symmetric criteria for the wave function $\widetilde{\Xi}$ with respect to $P_x\widetilde{S}$, namely, we classify the solutions fulfilling the condition $P_{x}\widetilde{S}\boldsymbol{\Xi}=\pm\boldsymbol{\Xi}$ as even ($\boldsymbol{\Xi}^{(e)}$ for $+$) and odd ($\boldsymbol{\Xi}^{(e)}$ for $-$). In this form, the coefficients of $\Xi_{\operatorname{II}}$ in~\eqref{Xi} are $\alpha_{\operatorname{II}}=\pm \beta_{\operatorname{II}}$ for even ($+$) and odd ($-$) functions, so that after evaluating the boundary condition at $x=L$ one obtains relations to determine the energies of even and odd states as
\begin{equation}
\tan\left(\frac{k_{\operatorname{II}}L}{2}\right)=F(E),  \quad
-\operatorname{cot}\left(\frac{k_{\operatorname{II}}L}{2}\right)=F(E), \quad
F(E)=\frac{E}{U_{0}-E}\sqrt{\frac{(E-U_{0})^{2}-m^{2}}{m^{2}-E^{2}}} ,
\label{trans-0}
\end{equation}
respectively, with $k_{\operatorname{II}}=\sqrt{(E-U_{0})^{2}-m^{2}}/\hbar v_{1}$.

Although the exact values of $E$ cannot be analytically determined for arbitrary $L$, one can still determine the exact number of even ($N^{(\operatorname{e})}$) and odd ($N^{(\operatorname{o})}$) bound states. The thorough analysis (see App.~\ref{sec:even-odd} for a detailed proof) leads to 
\begin{equation}
\begin{aligned}
& N^{(\operatorname{e})}=\left\lfloor \frac{L}{\pi\hbar v_{1}}\sqrt{\frac{U_{0}}{2}\left(m+\frac{U_{0}}{2} \right)}+\frac{1}{2}\right\rfloor - \left\lfloor \frac{L}{\pi\hbar v_{1}}\sqrt{\frac{U_{0}}{2}\left(-m+\frac{U_{0}}{2} \right)}+\frac{1}{2}\right\rfloor + 1 , \\
& N^{(\operatorname{o})}=\left\lfloor \frac{L}{\pi\hbar v_{1}}\sqrt{\frac{U_{0}}{2}\left(m+\frac{U_{0}}{2} \right)}\right\rfloor - \left\lfloor \frac{L}{\pi\hbar v_{1}}\sqrt{\frac{U_{0}}{2}\left(-m+\frac{U_{0}}{2} \right)}\right\rfloor + 1 ,
\end{aligned}
\label{N-even-odd}
\end{equation}
with $\lfloor \cdot \rfloor$ the \textit{floor function}. 

From the latter, it is clear that at least one even and one odd bound state always exist, regardless of the potential width and strength. Particularly, for small enough $L\rightarrow 0$, one obtains the $E\rightarrow 0$ and odd $E\rightarrow -m$ as the even and odd bound state energies, respectively.

\begin{figure}
\centering
\subfloat[][$U_{0}=1.5$]{\includegraphics[width=0.3\textwidth]{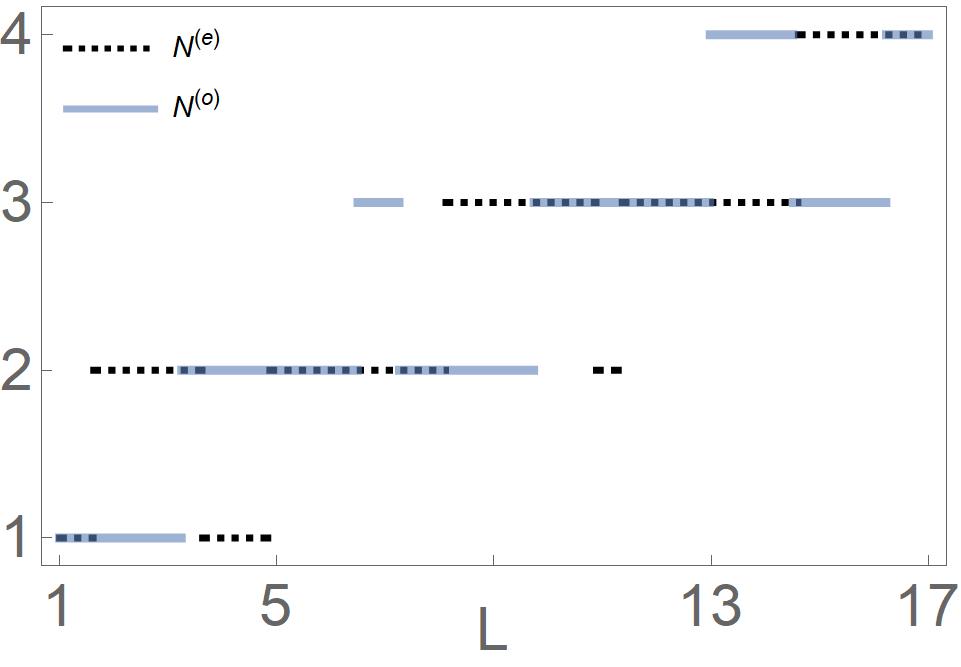}
\label{fig:N-even-odd}}
\hspace{2mm}
\subfloat[][]{\includegraphics[width=0.3\textwidth]{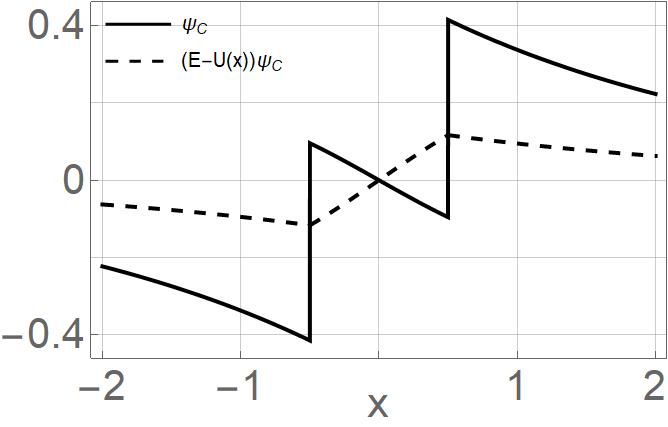}
\label{fig:psiC-even-odd}}
\hspace{2mm}
\subfloat[][]{\includegraphics[width=0.3\textwidth]{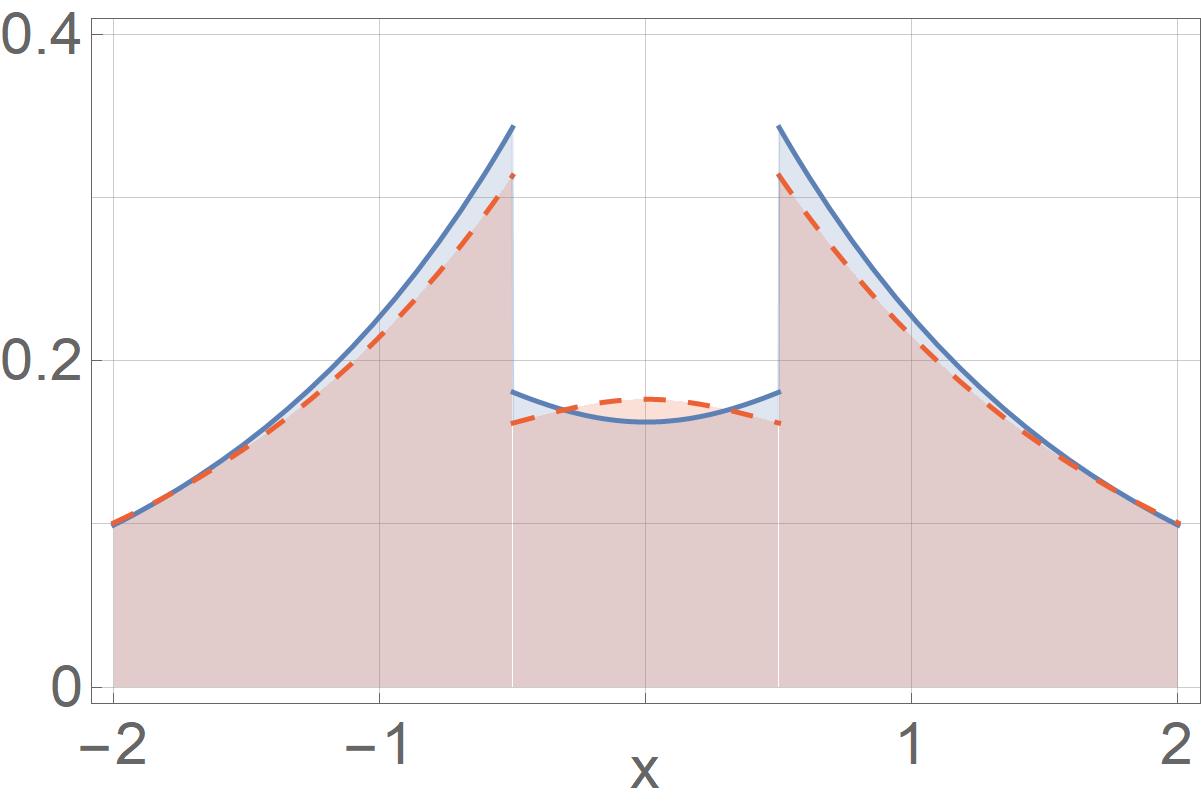}
\label{fig:PD-even-odd}}
\caption{(In units of $\hbar$=1) (a) Number of even (dotted) and odd (blue-thick) bound states as a function of $L$ for $v_{1}=1$, $m=0.5$ and $U_{0}=1.5$. (b) Eigensolution component $\psi_{C}$ and $(E-U(x))\psi_{C}$ for $L=v_{1}=1$ and $U_{0}=1.5$ and the even bound state energy $E\approx 0.281398$. (c) Probability distribution associated with the eigenvalues $E\approx 0.281398$ (blue-solid) and $E\approx -0.32653$ (red-dashing) and the same parameters as in (b).}
\end{figure}

Since the floor function is discontinuous, the number of bound states does not necessarily grow continuously for increasing values of $L$. That is, for $L=L_{0}$ with $N^{(\operatorname{e,o})}$ bound states, there might be a $L=L_{1}>L_{0}$ such that $(N^{(\operatorname{e,o})}-1)$ are generated. This is indeed depicted in Fig.~\ref{fig:N-even-odd} for fixed potential depth and different potential length $L$.

As discussed in Sec.~\ref{subsec:Spin1D}, the component $\psi_{C}$ might not be continuous, which can lead to discontinuous probability densities. Still, one may verify the validity of the bound state eigenvalues $E$ obtained from~\eqref{trans-0} by substituting it into $(E-U(x))\psi_{C}$, which should be a continuous function\footnote{It follows from $k_{y}\Psi_B+(U(x)-E)\Psi_C=0$, which the third of the coupled equations represented by (\ref{Dirac-H2}).}. Particularly, from Fig.~\ref{fig:N-even-odd}, one notices that $L=1$ and $U_{0}=1.5$ lead to one even ($E_{0}^{(\operatorname{e})}\approx 0.281398$) and one odd ($E_{0}^{(\operatorname{o})}\approx -0.32653$) bound state energy eigenvalue. The component $\psi_{C}$ and $(E-U(x))\psi_{C}$ are depicted in Fig.~\ref{fig:psiC-even-odd} for $E\approx 0.281398$, which verifies the required continuity condition for the latter function. The same conclusion is drawn for $E\approx -0.32653$. Furthermore, the corresponding probability distributions associated with $\widetilde{\boldsymbol{\Psi}}^{(\operatorname{e})}$ and $\widetilde{\boldsymbol{\Psi}}^{(\operatorname{o})}$ are depicted in Fig.~\ref{fig:PD-even-odd} in blue-solid and red-dashed, respectively, which are discontinuous.

\section{Scattering states and transmission amplitudes}
\label{subsec:U-scattering}
Let us now focus on the scattering of the plane waves on the barrier and the related phenomena. This is obtained when plane-wave-like solutions are present in the regions I and III, which corresponds to the eigenvalues $E\in(-\infty,-\widetilde{m})\cup(\widetilde{m},\infty)$. Without loss of generality, we consider only outgoing waves in region III and outgoing together with incoming waves in region I. The coefficients of the wave function (\ref{Xi}) are then fixed in the following manner, 
\begin{equation}
\alpha_{\operatorname{I}}=1,\quad \beta_{\operatorname{I}}=r,\quad \alpha_{\operatorname{III}}=t,\quad \beta_{\operatorname{III}}=0.
\end{equation}
The complex constants $t$ and $r$ can be calculated from (\ref{transfermatrix}) as
\begin{equation}\label{rt}
t=\frac{1}{m_{22}},\quad r=-\frac{m_{21}}{m_{22}}.
\end{equation}
The coefficients $r$ and $t$ define the reflection and transmission coefficients $R=|r|^2$ and $T=|t|^2$ that satisfy $R+T=1$. The later expression can be directly verified by substituting from (\ref{rt}) when taking into account that  there holds $m_{11}=m_{22}^*$ and $m_{12}=m_{21}^*$.
After some calculations, one obtains,
\begin{equation}
r=\sin(k_{\operatorname{II}} L)\frac{-2A(B-B')+i\left(A'^2-A^{2}+(B-B')^{2}\right)}{2AA'\cos(k_{\operatorname{II}}L)-i\sin(k_{\operatorname{II}}L)\left((B-B')^{2}+A^{2}+A'^{2}\right)},
\label{reflection1}
\end{equation}
where
\begin{equation}
\frac{A}{v_{1}}=\frac{E k_{\operatorname{I}}}{E^{2}-m^{2}}, \quad \frac{A'}{v_{1}}=\frac{(E-U_{0}) k_{\operatorname{II}}}{(E-U_{0})^{2}-m^{2}} , \quad \frac{B}{v_{2}}=\frac{mk_{y}}{E^{2}-m^{2}} , \quad \frac{B'}{v_{2}}=\frac{m k_{y}}{(E-U_{0})^{2}-m^{2}} ,
\end{equation}
and $k_{\operatorname{I}}=\frac{\sqrt{E^{2}-\widetilde{m}^{2}}}{\hbar v_{1}}$, $k_{\operatorname{II}}=\frac{\sqrt{(E-U_{0})^{2}-\widetilde{m}^{2}}}{\hbar v_{1}}$. 
This expression also holds in cases where solutions in the region II are evanescent waves. 

Eq.~\eqref{reflection1} is a handy expression to understand the transmission of incoming waves from the region I and traveling to the region III. As an immediate case, one may consider incident waves with energy $E=U_{0}$. This leads to $A'=0$ and the reflection coefficient becomes an unimodular number. That is, $r=e^{i\phi}$, with $\phi\equiv\phi(U_{0},m,k_{y})$ some phase, the exact form of which is irrelevant as the transmission coefficient becomes $T=1-\vert r\vert^{2}=0$. In other word, incoming waves with energy equal to the potential barrier height are reflected, up to a phase shift.

In this work, particular interest is paid to cases in which perfect tunneling exists, $T=1$.  Such a tunneling is obtained whenever $r=0$, which ensures that $t$ is a unimodular complex number, and incident and transmitted waves share their amplitude, but the latter carries a relative phase shift $t$ as a leftover of its interaction with the barrier. For the sake of clarity, we split our discussion in two cases.

\subsection*{Normal incidence ($k_{y}=0$)}
In this case, the reflection coefficient becomes simpler since $B=B'=0$, $k_{\operatorname{I}}=\sqrt{E^{2}-m^{2}}/\hbar v_{1}$, and $k_{\operatorname{II}}=\sqrt{(E-U_{0})^{2}-m^{2}}/\hbar v_{1}$. The numerator in $r$ becomes proportional to $m\sin(k_{\operatorname{II}}L)$. Therefore, for the gapless lattice setup ($m=0$), perfect tunneling occurs for any arbitrary energies in $E\in(-\infty,-m)\cup(m,\infty)$. This effect was reported in graphene~\cite{Kat06,All11,Jak11} and pseudospin-1 lattices~\cite{Ill17,Urb11}.
    
For $m\neq 0$, perfect tunneling does exist for specific energies so that $k_{\operatorname{II}}L=n\pi$, with $n=1,\ldots$. The exact resonant energies are straightforward to compute and are presented in a more general case below. However, it is worth to analyze the behavior of $T=1-\vert r\vert^{2}$ when the barrier height is large enough, $U_{0}\gg m,E$, for fixed and finite $E$. The straightforward calculations show that
\begin{equation}
T\approx \frac{1}{1+\frac{m^{4}}{4E^{2}(E^{2}-m^{2})}\sin^{2}\left(\frac{U_{0}L}{\hbar v_{1}}\right)} .
\end{equation}
It reveals that, despite the lack of the perfect tunneling, the transmission  converges to a non-null value as the electrostatic barrier increases indefinitely. This is known as Klein paradox~\cite{Dom99}, and it is in sharp contrast with the non-relativistic case, where transmission becomes smaller for larger barrier heights. 

\subsection*{Oblique incidence ($k_y\neq 0$)}
$\bullet$ \textbf{Super-Klein tunneling} When $B=B'$ and $A=\pm A'$ in from~\eqref{reflection1}, the reflection coefficient vanishes and the transmission becomes perfect ($T=1$). This is achieved when $E=U_{0}/2$. One thus has perfect tunneling regardless of the incidence angle for $E=U_{0}/2$.
This phenomenon is called the \textit{super-Klein tunneling}, already reported for pseudospin-1 lattice models with gapless dispersion and flat bands\cite{Urb11,Xu17,Bet17}, as well as in pseudospin-1/2 graphene lattices~\cite{Con20}. Here, we note that the presence of the mass term ($m\neq 0$) does not break the super-Klein tunneling as long as $U_{0}>2m$. However, super-Klein tunneling is altogether lost by tuning the electrostatic barrier such that $0<U_{0}<2m$, as no plane-wave solutions exist for $E=U_{0}/2$. This highlights the effects of the mass term (band-gap) on the transmission properties.

\begin{figure}
\centering
\subfloat[][]{\includegraphics[width=0.27\textwidth]{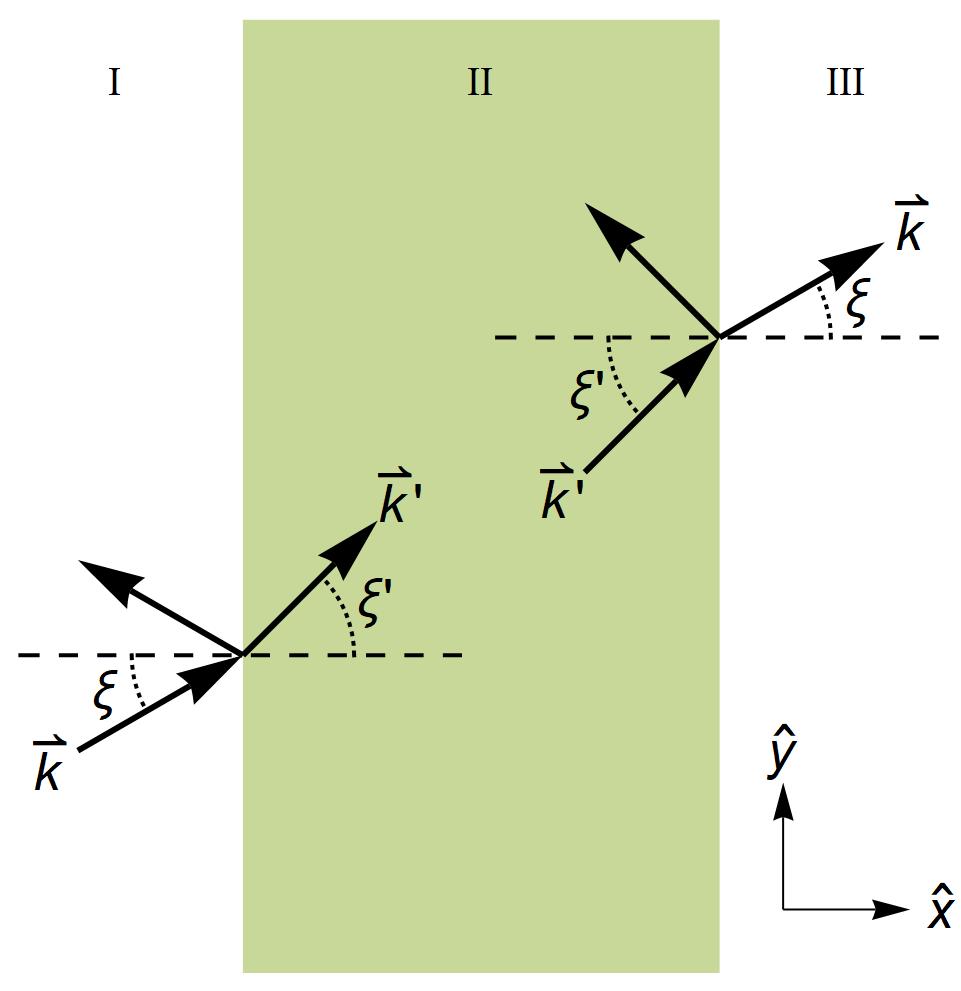}
\label{fig:barrier1}}
\hspace{2mm}
\subfloat[][]{\includegraphics[width=0.3\textwidth]{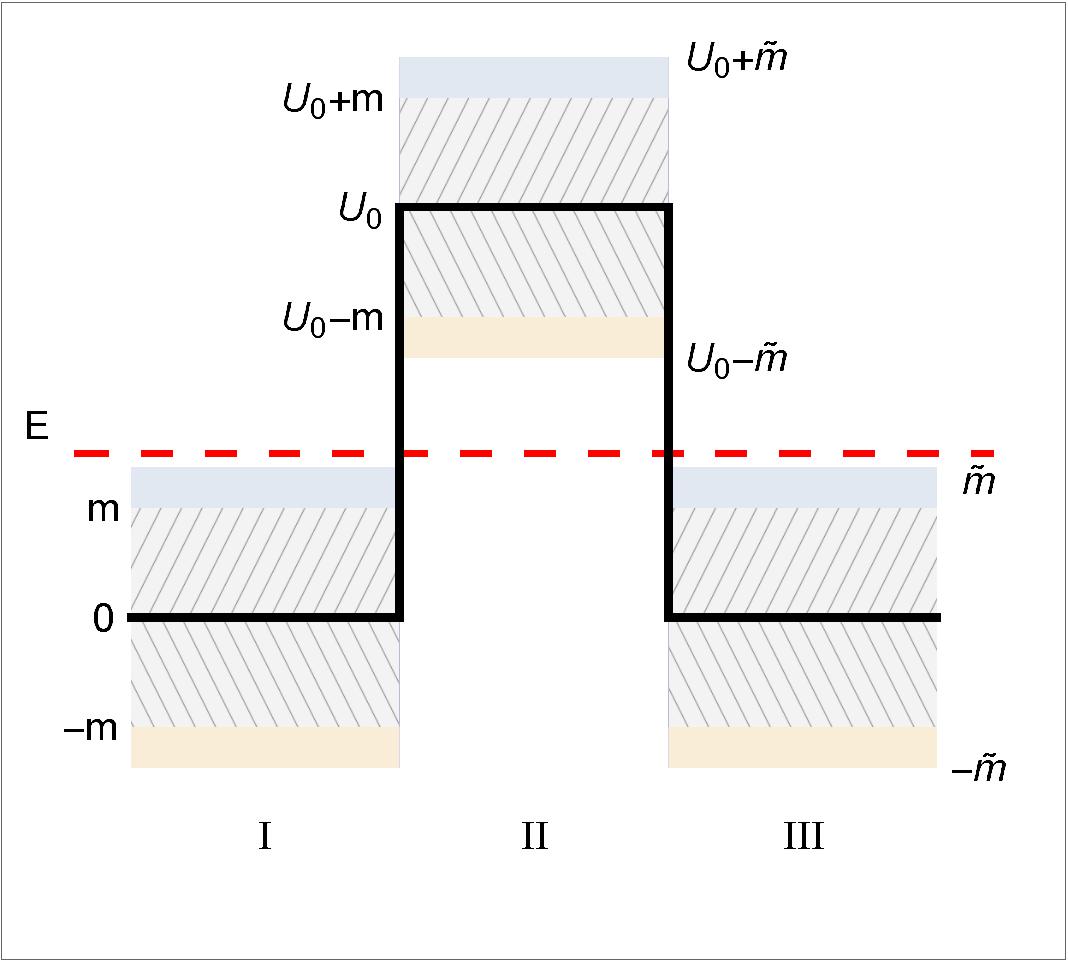}
\label{fig:scattering-0}}
\hspace{2mm}
\subfloat[][]{\includegraphics[width=0.3\textwidth]{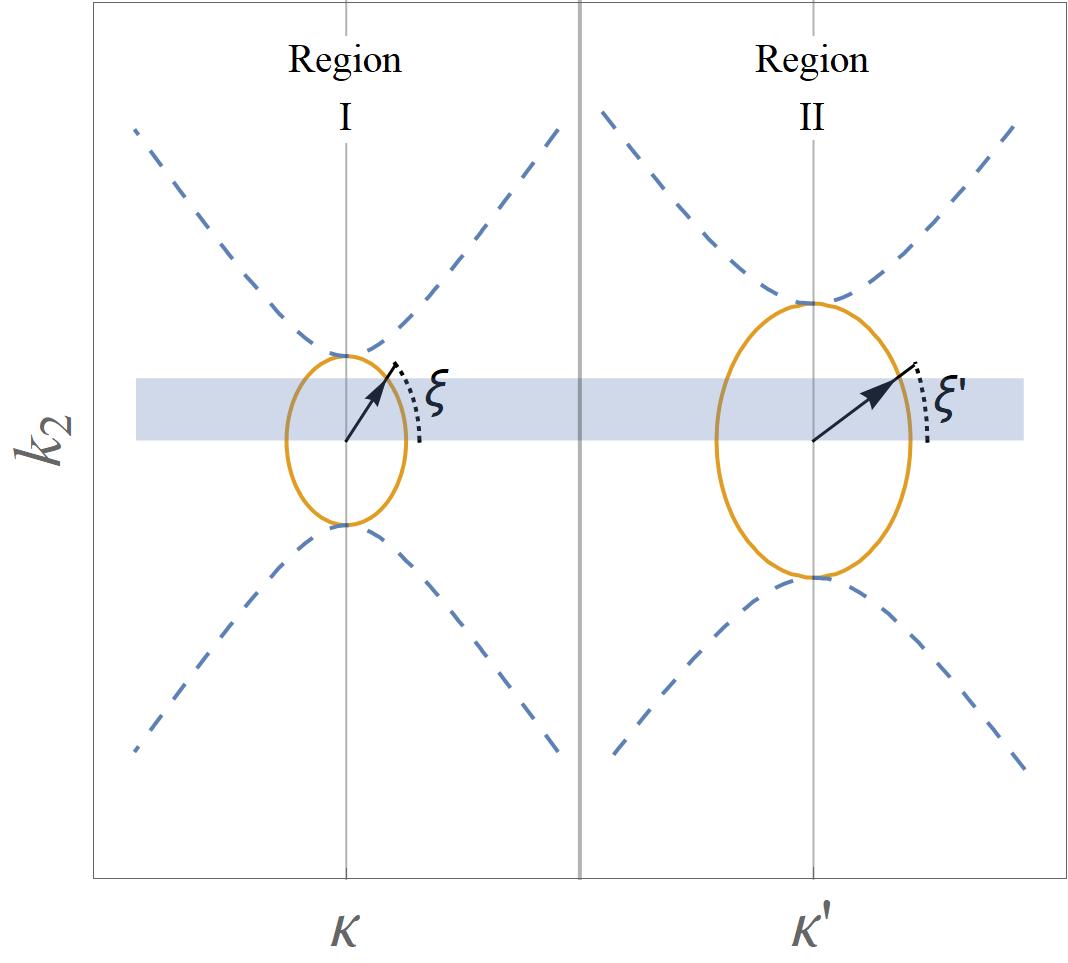}
\label{fig:scattering-1}}
\caption{(b) Scattering configuration (upper-view) for an incident wave $\vec{k}$ (region I), with incident angle $\xi$, traveling through an electrostatic barrier (green-shaded area). The wave refracts into region II as a wave with vector $\vec{k}'$ and transmitted angle $\xi'$. (b) Energy configuration of the panel (a) with an incident wave with energy $E>\widetilde{m}$ (red-dashed line). (c) Energy curves spanned by $\kappa,k_{y}$ (region I and III) and $\kappa',k_{y}$ (region II) for $E>\widetilde{m}$ fixed as in panel (b).}
\end{figure}

\noindent
$\bullet$ \textbf{Generalized Snell-Descartes law}
It is convenient to define the two-dimensional momentum vectors $\vec{k}=(k_{\operatorname{I}},k_{y})$ and $\vec{k}'=(k_{\operatorname{II}},k_{y})$ that characterize the incident wave and the wave traveling through the electric barrier, respectively. The incident and transmitted angles are defined as $\xi=\arctan(k_{y}/k_{\operatorname{I}})$ and $\xi'=\arctan(k_{y}/k_{\operatorname{II}})$, respectively, see Fig.~\ref{fig:barrier1}. Contrary to the bound state case of Sec.~\ref{subsec:U-confinement}, plane-wave solutions only exist in the region I for bounded values of $k_{y}$, i.e., $\vert k_{y}\vert<k_{y;c}=\sqrt{E^{2}-m^{2}}/\hbar v_{2}$. This alternatively implies that scattering phenomenon is available for restricted values of the effective-mass term $\widetilde{m}$. This is depicted in Fig.~\ref{fig:scattering-0}, from which it is also clear that, for $\vert k_{y}\vert > k_{y;c}$, the shaded area covered by the effective-mass region overlaps with the energy $E$, leading to evanescent-wave solutions in the region I.

From the dispersion relations in the regions I and II, together with the fact that $k_{y}$ is constant across all regions, one can establish a relation between the incident and transmitted angles $
\xi$ and $\xi'$ of Fig.~\ref{fig:barrier1},
\begin{equation}
\frac{\operatorname{tan}\xi'}{\operatorname{tan}\xi}\sqrt{\frac{v_{1}^{2}+v_{2}^{2}\operatorname{tan}^{2}\xi}{v_{1}^{2}+v_{2}^{2}\operatorname{tan}^{2}\xi'}}=\sqrt{\frac{E^{2}-m^{2}}{(E-U_{0})^{2}-m^{2}}}.
\label{mod-snell}
\end{equation}
For $v_{1}=v_{2}$, one recovers the same Snell-Descartes law previously reported for graphene~\cite{All11}, and to the Snell's law obtained for pseudospin-1 lattices with $m=0$ reported in~\cite{Bet17}.

Since we are considering $U_{0}>2m$, we get the following information about the transmitted angle:
\begin{itemize}
\item For $E\in\left( m,\frac{U_{0}}{2} \right)$, there exists a transmitted angle $\xi'$ for every incident angle $\xi\in(-\pi/2,\pi/2)$.
\item For $E=\frac{U_{0}}{2}$, the transmitted and incident angles are equal, $\xi'=\xi$. 
\item For $E\in\left( \frac{U_{0}}{2},U_{0}-m \right)\cup (U_{0}+m,\infty)$, there are transmitted angles $\xi'\in(-\pi/2,\pi/2)$ only for $\xi\in(-\xi_{c},\xi_{c})$, with the critical angle $\tan^{2}\xi_{c}=\frac{v_{1}^{2}}{v_{2}^{2}}\frac{(E-U_{0})^{2}-m^{2}}{2 U_{0}\left(E-\frac{U_{0}}{2} \right)}$. For other values of $\xi$, the solutions in the region II are evanescent waves.
\item For $E\in(U_{0}-m,U_{0}+m)$, there are only evanescent waves in the region II. 
\end{itemize}

\noindent
$\bullet$ \textbf{Fabry-P\'erot resonances} Perfect transmission occurs for other energies as well, nevertheless, it gets angle dependent. The reflection coefficient (\ref{reflection1}) vanishes for $k_{\operatorname{II}}L =n\pi$, with $n\in\mathbb{Z}^{+}$. Since $k_{\operatorname{II}}$ is in turn a function of the incidence angle $\xi$ and the energy $E$, one may conclude that, for a fixed energy $E$, perfect reflection appears only for some specific incidence angles. These are usually known as tunneling resonances or \textit{Fabry-P\'erot} resonances~\cite{All11}, and are given as a function of the incident angles $\xi$ as
\begin{equation}
E^{(\operatorname{res})}_{\pm;n}=\left(1+\frac{v_2^2}{v_1^2}\tan^2\xi\right) \left(U_{0}\pm\sqrt{U_0^2-\frac{1}{1+\frac{v_{2}^{2}}{v_{1}^{2}}\tan^{2}\xi}\left(U_{0}^{2}-\frac{\hbar^{2}\pi^{2}v_{1}^{2}(n+1)^{2}}{L^{2}}-\frac{m^{2}}{1+\frac{v_{2}^{2}}{v_{1}^{2}}\tan^{2}\xi}\right)}\right) ,
\label{resonance}
\end{equation}
with $n=0,1,\ldots$. It is then said that $E_{\pm;n}^{(\operatorname{red})}$ is a resonant energy provided that $E_{\pm;n}^{(\operatorname{red})}\in(-\infty,-\widetilde{m})\cup(\widetilde{m},\infty)$.

\begin{figure}
\centering
\subfloat[][]{\includegraphics[width=0.35\textwidth]{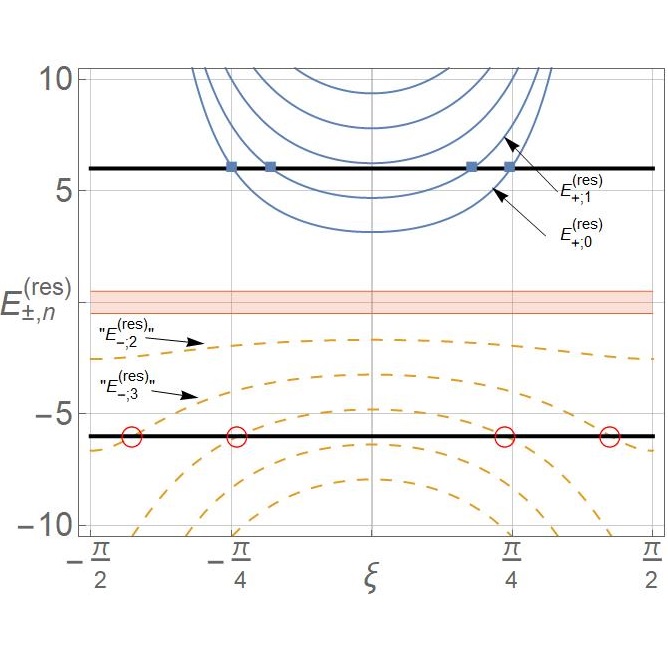}
\label{fig:resonance-1}}
\hspace{5mm}
\subfloat[][]{\includegraphics[width=0.2\textwidth]{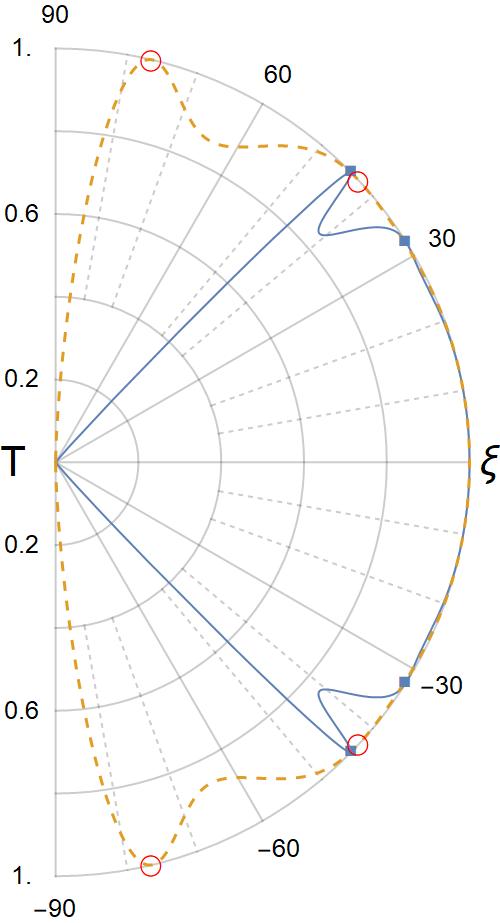}
\label{fig:resonance-0}}
\caption{(Units of $\hbar=1$) (a) Tunneling resonance energies $E_{+;n_{+}}$ (blue-solid) and $E_{-;n_{-}}$ (orange-dashed) for $n_{+}=0,1,2,3,4$ and $n_{-}=2,3,4,5$ as a function of $\xi(-\pi/2,\pi/2)$. The shaded area denotes the band-gap, where evanescent waves exist in the region I. The filled-squares and open-circles markers denote the angles at which resonance exists for $E=6$ and $E=-6$, respectively. The parameters have been fixed as $v_{1}=v_{2}=1$, $L=2$, $m=0.5$ and $U_{0}=1.5$. (b) Transmission amplitude $T$ as a function of the incident angle $\xi$ for $E=6$ (blue-solid) and $E=-6$ (orange-dashed) with the same parameters as in panel (a).
}
\label{fig:resonance}
\end{figure}

These resonant energies behave asymptotically as $\lim_{\xi\rightarrow\pm\pi/2}E_{+;n}^{(\operatorname{res})}\rightarrow\infty$ and $\lim_{\xi\rightarrow\pm\pi/2}E_{-;n}^{(\operatorname{res})}\rightarrow(2U_{0})^{-1}\left( U_{0}^{2}-\pi^{2}\frac{v_{1}^{2}(n+1)^{2}}{L^{2}} \right)$. Thus, for almost perpendicular incident waves ($\xi\sim \pm\pi/2$), one requires larger and larger energies in order to recover the resonances at $E_{+;n}^{(\operatorname{res})}$, whereas finite and well-defined energy values are required for the resonances $E_{-;n}^{(\operatorname{res})}$. This behavior is depicted in Fig.~\ref{fig:resonance-1}, whereas the transmission amplitude is depicted in Fig.~\ref{fig:resonance-0} as a function of the incident angle $\xi$.

\section{Remarks on the flat-band solutions}
The piece-wise continuous nature of the electrostatic interaction~\eqref{U-pot} allows the generation of two flat band energies, one located at $E=U_{0}$ for the region II, and another one at $E=0$ for the regions I and III. Although the boundary conditions are the same in both cases, the allowed matching solutions have a different behavior.

Let us first consider $E=U_{0}$ and $k_{y}$ so that plane-wave solutions exist for the regions I and III. For generality, we consider incoming and outgoing plane waves in regions I and III, and a general flat-band solution in II. Here, the waves entering the interaction zone from the left and right have an amplitude $I_{1}$ and $I_{2}$, respectively, with $I_{1,2}\in\mathbb{R}$. Additionally, we fix $I_{1}^{2}+I_{2}^{2}=1$. Under these considerations, we have the general solutions
\begin{equation}
\boldsymbol{\Xi}=
\begin{cases}
I_{1}\boldsymbol{\Xi}_{k_{\operatorname{I}}}+A_{1}\boldsymbol{\Xi}_{-k_{\operatorname{I}}} \quad &x<-L/2 \\
\boldsymbol{\Xi}_{fb} \quad &\vert x\vert<L/2 \\
I_{2}\boldsymbol{\Xi}_{k_{\operatorname{I}}}+A_{2}\boldsymbol{\Xi}_{-k_{\operatorname{I}}} \quad &x>L/2 
\end{cases}
\label{fb-trasmission}
\end{equation}
where $A_{1,2}\in\mathbb{C}$, $\boldsymbol{\Xi}_{fb}=(m\chi,i\hbar v_{2}k_{y}\chi,-\hbar v_{1}\chi')^{T}$, with $\chi$ a complex-valued function, and $\boldsymbol{\Xi}_{\pm k_{\operatorname{I}}}$ the solutions \eqref{fp-gen} evaluated at $E=U_{0}$. By imposing the boundary conditions~\eqref{cont}, one obtains the relations $A_{1}=I_{1}e^{-2i\phi}e^{ik_{\operatorname{I}}L}$ and $A_{2}=I_{2}e^{2i\phi}e^{-ik_{\operatorname{I}}L}$, with $\phi=\arctan(v_{2}k_{y}U_{0}/mv_{1}k_{\operatorname{I}})$, whereas the arbitrary function $\chi$ is restricted to fulfill the following relations at the boundaries,
\begin{equation}
\chi\left(-\tfrac{L}{2}\right)=\frac{2v_{1}k_{\operatorname{I}}I_{1}e^{-i(\phi-k_{\operatorname{I}L})}}{\sqrt{m^{2}v_{1}k_{\operatorname{I}}^{2}+v_{2}k_{y}^{2}U_{0}^{2}}}, \quad \chi\left(\tfrac{L}{2}\right)=-\frac{2v_{1}k_{\operatorname{I}}I_{2}e^{i(\phi-k_{\operatorname{I}L})}}{\sqrt{m^{2}v_{1}k_{\operatorname{I}}^{2}+v_{2}k_{y}^{2}U_{0}^{2}}}.
\label{chi1}
\end{equation}
Given the arbitrary nature of $\chi$, one may alternatively rewrite it as $\chi=\tfrac{2v_{1}k_{\operatorname{I}}e^{i(\phi-k_{\operatorname{I}L})2x/L}}{\sqrt{m^{2}v_{1}k_{\operatorname{I}}^{2}+v_{2}k_{y}^{2}U_{0}^{2}}}\widetilde{\chi}$, where $\widetilde{\chi}(-L/2)=I_{1}$ and $\widetilde{\chi}(L/2)=-I_{2}$. 

Thus, the coupling of incident waves to the flat-band solution leads to a scattering problem in which the waves entering the interaction region are completely reflected inside their respective regions. Still, the flat band solutions allowed during such a process must fulfill the boundary conditions~\eqref{chi1}. Note that one also has the conservation property $\vert A_{1}\vert^{2}+\vert A_{2}\vert^{2}=I_{1}^{2}+I_{2}^{2}=1$. The latter results hold whenever waves enter from only one region, say $I_{1}=1$ and $I_{2}=0$. In such a case, we have a perfect reflection in region I, up to a phase in the reflected wave. 

Flat-band solutions also occur for $E=0$ in regions I and III. The arbitrary nature of the solutions in those flat bands can be tuned so that finite-norm solutions appear. The corresponding wave function in region II can be found using the boundary conditions, and the calculations are as straightforward as the scattering case presented above. 

\section{Concluding remarks}
\label{sec:conclusions}
In this manuscript, it was shown that the existence of a rectangular electrostatic barrier always produces at least two bound states for $k_{y}=0$, and generates more bound states at different energies for increasing values of $k_{y}$. Interestingly, it was found that even for the asymptotic values $\hbar v_{2} k_{y}\rightarrow\infty$, the associated current density parallel to the barrier is bounded by $\pm\hbar v_{2}$, where $v_{2}=2at_{2}$. Thus, the current is linear on the hopping amplitude across the $\hat{y}$-direction, as expected. 

It is worth remarking that dispersion relations obtained from~\eqref{trans} identify the energies for which electrons localize in the $x$-direction, and propagation is allowed in the $\hat{y}$-direction is still possible. However, by exploiting the separability of free-particle solutions, one can always construct linear combinations so that electrons localize in the $\hat{y}$-direction as well. Such a procedure has been discussed in~\cite{Jak16} for graphene. For instance, in the example provided in Fig.~\ref{fig:bound-k2-E}, one can take the energies associated with blue-solid and red-dashed curves as they exist for any $k_{y}\in\mathbb{R}$. From the relations~\eqref{N-even-odd}, one can ensure that at least two of such dispersion relations always exist. Additional caution must be taken for the other dispersion relations, as they only exist for intervals $k_{y}\in\mathcal{S}\subseteq\mathbb{R}$, and the linear combination must be constructed accordingly to that interval. Devising such packages is a task beyond the scope of the current work and will be discussed elsewhere, as it deserves attention by itself.

On the one hand, for the scattering-wave regime, we have proved that even in the gapped case ($m\neq 0$), the Lieb lattice supports super-Klein tunneling for an energy equal to half of the electric barrier, $E=U_{0}/2$, provided that $U_{0}>2m$. For the gapless case, we recover the same results previously reported for gapless $T_{3}$ lattices~\cite{Urb11} and ultra-cold atoms trapped in optical lattices~\cite{Shen10}. On the other hand, we identified a new modified Snell-like law valid for anisotropic Fermi velocities $v_{1}\neq v_{2}$. The latter allows us to identify the Fabry-P\'erot resonant transmission, which defines a relation between the incident energy and the incident-wave angle required to produce perfect tunneling up to a phase factor. Interestingly, for negative energies, perfect transmission is achievable for finite energies at incident waves almost perpendicular to the barrier. This is not the case for positive energies, as it is shown that the required energies diverge.

The existence of flat-band solutions poses an additional case not available in graphene lattices. The latter allows the coupling of the solutions and determining the transmission properties, which in this case, leads to perfectly reflected waves. Since the flat-band solutions are defined in terms of degenerate Bloch waves, there is an infinite family of solutions that allows such a reflection, as long as they fulfill the boundary condition~\eqref{chi1}. 


\section*{Acknowledgments}
K.Z. acknowledges the support from the project ``Physicists on the move II'' (KINE\'O II) funded by the \textit{Ministry of Education, Youth, and Sports} of the Czech Republic, Grant No. CZ.02.2.69/0.0/0.0/18 053/0017163. 

\appendix
\setcounter{section}{0}
\section{Determining the number of even and odd bound states}
\label{sec:even-odd}
\renewcommand{\thesection}{A-\arabic{section}}
\renewcommand{\theequation}{A-\arabic{equation}}
\setcounter{equation}{0}  

In this appendix, we present the derivation of the number of bound states for even bound states presented in~\eqref{N-even-odd}. The procedure applies straightforwardly to the odd case as well. It is convenient to define the intervals
\begin{equation}
I_{0}=\left(0,\frac{\pi}{2}\right), \quad I_{n}=\left(\frac{\pi}{2}+(n-1)\pi,\frac{\pi}{2}+n\pi\right), \quad n=1,2,\ldots ,
\label{intervals}
\end{equation}
so that $\mbox{tan(x)}$ is nonsingular for $x\in I_{n}$. Furthermore, if $x\in\cup_{k=p_{1}}^{k=p_{2}}I_{k}$, then $\mbox{tan(x)}$ has $(p_{2}-p_{1})$ singularities.  

To determine the number of even bound states, one must find the number of interceptions of $F(E)$ in~\eqref{trans-0} and the periodic function $\tan(\tfrac{k_{\operatorname{II}}L}{2})$ in the interval $E\in(-m,m)$. To this end, one may notice that $F(E)$ is a monotonously increasing function of $E\in(-m,m)$ that tends to $\mp\infty$ for $E\rightarrow\mp m$, and vanishes for $E=0$. The latter means that $F(E)$ defines the bijection $F(E):(-m,m)\mapsto\mathbb{R}$. On the other hand, $\partial k_{\operatorname{II}}/\partial E<0$ for $E\in(-m,m)$, and one thus concludes that $\tan(k_{\operatorname{II}}L/2)$ (and also $-\operatorname{cot}(k_{\operatorname{II}}L/2)$) is a monotonously decreasing function of $E$ in each of the intervals $\tfrac{k_{\operatorname{II}}L}{2}\in I_{n}$, with $n=0,1,\ldots$. This property, combined with the fact that $F(E):(-m,m)\mapsto\mathbb{R}$ is a monotonously increasing function, one concludes that interception of both functions always exist. One must determine the exact number of interceptions.

By exploiting the fact that $\tan(k_{\operatorname{II}L/2})$ is a periodic function, one just needs to count the number of periods inside the interval $E\in(-m,m)$ for arbitrary $U_{0}$ and $L$, which is equal to the number of singularities plus one. $m$ is a lattice parameter, so it is assumed to be a fixed value. Be $\sigma_{\pm}:=\tfrac{k_{\operatorname{II}}L}{2}\vert_{E=\mp m}=\frac{L}{\hbar v_{1}}\sqrt{\frac{U_{0}}{2}\left(\pm m+\frac{U_{0}}{2}\right)}$ so that the domain of $\tan\left(\tfrac{k_{\operatorname{II}}L}{2}\right)$ lies in the interval $(\sigma_{-},\sigma_{+})$ for $E\in(-m,m)$. 

Now, if $\sigma_{-}\in I_{r_{1}}$ and $\sigma_{+}\in I_{r_{2}}$, with $r_{2}>r_{1}$ and $r_{1,2}=0,1,\ldots$, then, $\tan(k_{\operatorname{II}}L/2)$ has $r_{2}-r_{1}$ singularities for $E\in(-m,m)$ and intercepts $F(E)$ exactly $(r_{2}-r_{1}+1)$-times. That is, $N^{(\operatorname{e})}=r_{2}-r_{1}+1$. The values of $r_{1,2}$ are found by exploiting the fact that $\lfloor \tfrac{x}{\pi}+\tfrac{1}{2}\rfloor=r$ for $x\in I_{r}$. One thus has $\lfloor \tfrac{\sigma_{-}}{\pi}+\tfrac{1}{2} \rfloor=r_{1}$ and $\lfloor \tfrac{\sigma_{+}}{\pi}+\tfrac{1}{2} \rfloor=r_{2}$, which leads to the expression presented in~\eqref{N-even-odd}.

The same procedure applies to the odd solutions, where we define
the intervals $\widetilde{I}_{n}=(n\pi,(n+1)\pi)$, with $n=0,1,\ldots$, so that $\cot(x)$ is nonsingular for $x\in \widetilde{I}_{n}$. Since $-\cot(k_{\operatorname{II}}L/2)$ is monotonously decreasing for $k_{\operatorname{II}}L/2\in \widetilde{I}_{n}$, the same same reasoning used in the even case applies to the odd case, and one obtains $N^{(\operatorname{o})}$ in~\eqref{N-even-odd}.


\end{document}